\documentclass[12pt]{article}

\usepackage{amssymb}

\usepackage{latexsym}
\usepackage{epsfig}

\usepackage{tcolorbox}
\usepackage{tabularx} 
\usepackage{amsmath}  
\usepackage{graphicx} 
\usepackage{float}

\usepackage{cite} 
\usepackage[final]{hyperref} 

\usepackage{ulem}	
	
\usepackage{fancyhdr}
\usepackage{datetime}

\graphicspath{{images/}}

\usepackage{tcolorbox}
\usepackage{color,soul}

\definecolor{azure(colorwheel)}{rgb}{0.0, 0.5, 1.0}
\definecolor{DarkViolet}{RGB}{148,0,211}
\definecolor{MyDarkBlue}{rgb}{0,0.1,0.7}
\definecolor{DarkBlue}{RGB}{0,0,153}
\definecolor{amber}{rgb}{1.0, 0.49, 0.0}
\definecolor{amaranth}{rgb}{0.9, 0.17, 0.31}
\definecolor{nicered}{rgb}{0.7,0.1,0.1}
\definecolor{brown}{rgb}{0.5,0.1,0.1}
\definecolor{nicegreen}{rgb}{0.0,0.3,0.0}
\definecolor{tealgreen}{rgb}{0.0, 0.51, 0.5}

\newcommand{\newc}{\newcommand}
\newc{\com}[1]{\textcolor{amaranth}{#1}} 
\newc{\bako}[1]{\textcolor{DarkViolet}{#1}} 
\newc{\red}[1]{\textcolor{red}{#1}} 
	
\hypersetup{colorlinks,bookmarksopen,
bookmarksnumbered,
citecolor={nicered},
linkcolor={MyDarkBlue},
urlcolor={tealgreen},
pdfstartview=FitH}

\newcommand{\pd}{\partial}

\newcommand{\mc}{\mathcal}
\newcommand{\tx}{\mathrm}
\newcommand{\A}{{\cal{A}}}
\newcommand{\B}{{\cal{B}}}
\newcommand{\G}{{\mathcal{G}}}

\relax
\renewcommand{\theequation}{\arabic{section}.\arabic{equation}}
\def\be{\begin{equation}}
\def\ee{\end{equation}}
\def\bea{\begin{eqnarray}}
\def\eea{\end{eqnarray}}

\newcommand\fverb{\setbox\pippobox=\hbox\bgroup\verb}
\newcommand\fverbdo{\egroup\medskip\noindent%
                        \fbox{\unhbox\pippobox}\ }
\newcommand\fverbit{\egroup\item[\fbox{\unhbox\pippobox}]}

\newcommand{\bear}{\begin{eqnarray}}

\newcommand{\eear}{\end{eqnarray}}
\newbox\pippobox

\def\lab{\label}
\def\a{\alpha}
\def\b{\beta}
\def\g{\gamma}
\def\d{\delta}

\def\L{\Lambda}

\def\f{\phi}

\def\sq
\def\y{\psi}

\usepackage{orcidlink}
\def\idbako{\orcidlink{0000-0002-3012-6144}}
\def\idchar{\orcidlink{0000-0002-5364-4753}}
\def\idkand{\orcidlink{0000-0002-3018-5558}}
\def\idnico{\orcidlink{0000-0001-8686-4093}}

\begin{document}

\begin{titlepage}

\vspace*{1cm}
\begin{center}
{\bf \Large Compact objects of spherical symmetry in beyond Horndeski theories}
\bigskip \bigskip \medskip

{\bf A. Bakopoulos\idbako}$^{\,(a)}$\,\footnote{Email: a.bakop@uoi.gr} 
{\bf C. Charmousis\idchar}$^{\,(b)}$\,\footnote{Email: christos.charmousis@ijclab.in2p3.fr} 
{\bf P. Kanti\idkand}$^{\,(c)}$\,\footnote{Email: pkanti@uoi.gr} {\bf N. Lecoeur\idnico}$^{\,(b)}$\,\footnote{Email: nicolas.lecoeur@ijclab.in2p3.fr}

 \bigskip
 
$^{(a)}${\it Division of Applied Analysis, Department of Mathematics,\\
University of Patras, Rio Patras GR-26504, Greece}

$^{(b)}${\it Université Paris-Saclay, CNRS/IN2P3, IJCLab, 91405 Orsay, France}

$^{(c)}${\it Division of Theoretical Physics, Department of Physics,\\
University of Ioannina, Ioannina GR-45110, Greece}

\bigskip \medskip

\today

\bigskip \medskip
{\bf Abstract}\\
\end{center}

\noindent 
We analyse in all generality beyond Horndeski theories of shift symmetry in a static and spherically symmetric spacetime. By introducing four auxiliary functions, we write the field equations in a particularly compact form. We show that assuming additionally parity symmetry renders the system directly integrable giving multiple families of black-hole solutions. These have typically an asymptotically-flat Reissner-Nordstrom behaviour, and emerge in the presence of a canonical kinetic term for the scalar field. In the absence of parity symmetry, we present a general method which allows us to integrate the field equations by choosing the form of only one coupling function and an auxiliary quantity. This method leads to asymptotically flat and AdS black hole solutions with differing properties. We finally discuss disformal transformations within this context as a means of obtaining wormhole and black hole solutions in different theories. 
\bigskip \medskip

\end{titlepage}

\setcounter{page}{1}

\def\g{\gamma}
\def\go{\g_{00}}
\def\gi{\g_{ii}}


\tableofcontents

\section{Introduction}

The impressive income of observational data for compact objects have initiated a revolutionary epoch in the field of gravitational physics. Observations range from gravitational wave signals of binary mergers of relatively few solar masses (eg. \cite{LIGOScientific:2017vwq}, \cite{LIGOScientific:2020zkf}, \cite{LIGOScientific:2021qlt}) to images of supermassive black holes generated from radio-telescope networks  \cite{EventHorizonTelescope:2020qrl} and trajectories of stars orbiting the Galactic center \cite{GRAVITY:2018ofz}, to name just a few. Early findings are compatible with GR, raising however a number of questions like the nature of the secondary object in  a certain binary \cite{LIGOScientific:2020zkf} and the existence or not of the expected GR mass gap between neutron stars and black holes (for a recent discussion, see \cite{Charmousis:2021npl} and references within). Observational data, evolving from discovery towards precision, will  permit us to probe additional gravitational parameters, eventually checking the validity of no hair theorems, star trajectory parameters and possibly discovering novel effects (for example distinguishing a wormhole throat versus an event horizon \cite{Damour:2007ap}, \cite{Cardoso:2016rao}, \cite{Bambi:2021qfo}).

Scalar tensor theories,  \cite{Horndeski:1974wa},\cite{Gleyzes:2014dya}, \cite{Langlois:2015cwa}, \cite{Crisostomi:2016czh}, provide a robust and measurable departure from GR and are therefore very interesting geometric modifications of GR. Since GR is not a UV complete theory scalar tensor are expected to play an important role as curvature effects become stronger. This is particularly true for the smaller mass compact objects, in binaries for example, as their radius of curvature can be quite large. There is additional motivation for scalar tensor theories coming from dark energy, although applicability of data is questionable on cosmological scales \cite{deRham:2018red}.  An adjacent motivation is that most modified gravity theories admit a well defined scalar tensor limit. A classic example is the Horndeski cubic galileon (see eg. \cite{Kobayashi:2019hrl}, \cite{Babichev:2016fbg}), which originates from the 5 dimensional DGP \cite{Dvali:2000hr} braneworld model at the decoupling limit \cite{Nicolis:2004qq} (first discovered in massive gravity \cite{Arkani-Hamed:2002bjr}). Note that this model \cite{Dvali:2000hr}, from the 4 dimensional perspective has an infinite number of degrees of freedom, although as perceived from 5 dimensions it is a rather simple gravitational setup. Scalar tensor theories are also directly obtained from higher dimensional Lovelock theory \cite{Lovelock:1971yv} (see for example \cite{Charmousis:2014mia}) to scalar tensor via Kaluza Klein compactification \cite{VanAcoleyen:2011mj}, \cite{Charmousis:2012dw} or again from effective string theory actions \cite{Fradkin:1984pq}, \cite{Sen:1985qt}, \cite{Callan:1985ia}, \cite{Gross:1986mw}, \cite{Callan:1986jb}, \cite{Metsaev:1987zx} and their black hole solutions (see for example \cite{Campbell:1990ai}, \cite{Kanti:1995vq}, \cite{Kanti:1997br},\cite{Chan:1995fr}, \cite{Charmousis:2009xr}). 

By now there is a number of analytic and numerical black hole solutions in scalar tensor theories most of which obtained upon breaking hypotheses to no hair theorems (see for example \cite{Hui:2012qt}, \cite{Sotiriou:2014pfa},
\cite{Herdeiro:2015waa},
\cite{Babichev:2016rlq}, \cite{Antoniou:2017acq}, \cite{Antoniou:2017hxj}, \cite{Lehebel:2018zga}, \cite{Bakopoulos:2018nui}, \cite{Bakopoulos:2020dfg}, \cite{Bakopoulos:2021dry}). Most analytic solutions found exist in shift and parity symmetric scalar tensor theories typically including the K-essence (or $G_2$) and $G_4$ galileons \eqref{eq:Hfr} (see for example \cite{Babichev:2016rlq}, \cite{Volkov:2016ehx}, \cite{Lehebel:2018zga} and references within). It is quite interesting that shift symmetry allows a large class of these to have linear time dependence first introduced in \cite{Babichev:2013cya} and generalised in \cite{Kobayashi:2014eva}. This is associated to the presence of shift symmetry and a time like Killing vector field for static metrics. It is found that black hole metrics are generically close or identical to GR black hole solutions as the scalar field paints the spacetime following a geodesic congruence \cite{Charmousis:2019vnf}. This construction extends even to the stealth Kerr solution for a particular DHOST theory \cite{Charmousis:2019vnf}. Disformal mappings starting with stealth GR metrics provide for the first time an analytic solution of a rotating black hole which alters Kerr spacetime while passing all actual observational constraints \cite{Anson:2020trg}, \cite{Anson:2021yli}, \cite{Anson:2021msx}. 

Once we give up on parity or shift symmetry it is far more involved to obtain analytic solutions. Yet there is no physical reason to admit these symmetries; on the contrary putting them aside allows us to explore quite different spacetime geometries from GR often with quite different characteristics. The first asymptotically flat analytic solution found \cite{Charmousis:2012dw} was one obtained from Kaluza-Klein reduction originating from a Lovelock black hole \cite{Dotti:2005rc}, \cite{Bogdanos:2009pc} with a non trivial horizon geometry. Given its higher order nature this solution did not have a mass term with Newtonian fall off. However, as noticed more recently, an intriguing singular limit \cite{Glavan:2019inb} of  Lovelock theory and a careful analysis of the resulting scalar tensor theory, done is several different ways, \cite{Lu:2020iav}, \cite{Hennigar:2020lsl}, \cite{Fernandes:2020nbq}, \cite{Fernandes:2021dsb} gave three very interesting 4-dimensional black hole solutions (see the review \cite{Fernandes:2022zrq}) with interesting phenomenology \cite{Clifton:2020xhc}, \cite{Charmousis:2021npl}. These solutions originate from theories with and without shift symmetry. Stationary metrics are even harder to obtain in the absence of these symmetries. Some important progress has been achieved for a conformally coupled scalar field for type D metrics \cite{Anabalon:2009qt} and a rotating singular solution \cite{Astorino:2014mda}.

What is missing in the literature is a unifying and general study of spherically symmetric solutions for a given wide class of scalar tensor theories. This is the task we will attempt in this paper putting together different techniques and approaches in order to elaborate methods and eventually discover new black hole and wormhole solutions which are static and of spherical symmetry.
Given the above discussion, we will also choose to focus on non stealth solutions. As such we will consider a time independent scalar field. In what follows we will consider shift symmetric Horndeski plus beyond Horndeski theory. The former  is parametrised by four functions $\{G_i: i=2,..,5\}=\{G_2,G_3,G_4,G_5\}$ functions of  $\varphi$ and its kinetic density $X=-\frac{1}{2}\pd_\mu\phi\,\pd^\mu\phi$ :
\be
\label{eq:Hfr}
S_\tx{H} = \displaystyle\int \tx{d}^4x \sqrt{-g} \,\left(\mc{L}_2+\mc{L}_3+\mc{L}_4+\mc{L}_5\right),
\ee
with
\begin{align}
\mc{L}_2 &=G_2(X) ,
\label{eq:L2fr}
\\
\mc{L}_3 &=-G_3(X) \,\Box \phi ,
\label{eq:L3fr}
\\
\mc{L}_4 &= G_4(X) R + G_{4X} \left[ (\Box \phi)^2 -\nabla_\mu\pd_\nu\phi \,\nabla^\mu\pd^\nu\phi\right] ,
\label{eq:L4fr}
\\
\begin{split}
\mc{L}_5 &= G_5(X) G_{\mu\nu}\nabla^\mu \pd^\nu \phi - \frac{1}{6}\, G_{5X} \big[ (\Box \phi)^3 - 3\,\Box \phi\, \nabla_\mu\pd_\nu\phi\,\nabla^\mu\pd^\nu\phi
\\
&\quad + 2\,\nabla_\mu\pd_\nu\phi\, \nabla^\nu\pd^\rho\phi\, \nabla_\rho\pd^\mu\phi \big].
\label{eq:L5fr}
\end{split}
\end{align}
The latter is given by two additional higher order terms,
\begin{align}
\mathcal{L}^{\rm bH}_4&=F_4(X)\varepsilon^{\mu\nu\rho\sigma}\,\varepsilon^{\alpha\beta\gamma}_{\,\,\,\,\,\,\,\,\,\,\,\sigma}\,\partial_\mu\phi\,\partial_\alpha\phi\,\nabla_\nu\partial_\beta\phi\,\nabla_\rho\partial_\gamma\phi,\\[3mm]
\mathcal{L}^{\rm bH}_5&=F_5(X)\varepsilon^{\mu\nu\rho\sigma}\,\varepsilon^{\alpha\beta\gamma\delta}\,\partial_\mu\phi\,\partial_\alpha\phi\,\nabla_\nu\partial_\beta\phi\,\nabla_\rho\partial_\gamma\phi\,\nabla_\sigma\partial_\delta\phi.
\end{align}
The beyond Horndeski terms parametrised by $F_4$ and $F_5$ are not independent. They are related so as to evade the appearance of a ghost degree of freedom \cite{Crisostomi:2016tcp}. 
This relation reads
\begin{equation}
    X G_{5X} F_4= 3 F_5 (G_4-2X G_{4X})\,. \label{BH}
\end{equation}

In the next section we will set-up the problem and we will define special variables which will collectively trace the effects of functionals in the theory. We will see that depending on these one can have for example homogeneous or non homogeneous black holes. We will write the field equations in a compact and largely solvable form grouping together classes of Horndeski and beyond Horndeski theories. Then in section 3 we will study the case of parity symmetry. This case will be shown to be integrable in a sense that will be made clear in the third section. We will give a method providing exact solutions focusing on physically interesting cases involving eg., a canonical kinetic term. We will argue that although wormhole solutions can be constructed with relative ease their throats would be only theory dependant and not mass dependant. In section 4 we will leave aside the parity hypothesis. Focusing on the properties of the solutions originating from higher dimensional Lovelock theories we will generalise these in two different classes. Then in section 5 we will discuss disformal transformations of given solutions obtaining black hole and wormhole geometries (for non parity symmetric theories). 

\setcounter{equation}{0}
\section{Beyond Horndeski field equations and their analysis}%
Throughout this article we will consider a spherically symmetric Anzatz,
\be
ds^2=-h(r) dt^2 + \frac{dr^2}{f(r)} +r^2 d\Omega^2\,,
\ee
with a static scalar $\phi=\phi(r)$.  The independent field equations for this symmetry can be obtained in all generality. They first appeared in the Appendix of \cite{Lehebel:2018zga} and later presented in \cite{Bakopoulos:2021liw} with minor corrections.

It is particularly useful to write down the field equations for spherical symmetry and for zero scalar charge in all generality by introducing auxiliary functions $Z$ and $Y$ which are specific combinations of the theory functions.  Indeed we have,
\begin{align}
&Z(X)=2 X G_{4X}-G_4+ 4 X^2 F_4\,, \label{zdef}\\
&Y(X)=\frac{1}{2}(-2X)^{3/2}G_{5X}+3 (-2X)^{5/2} F_5\,. \label{ydef}
\end{align}
These functions ``include" the beyond Horndeski contribution to the relevant Horndeski terms. $Z$ and $Y$ will appear quite naturally when we discuss disformal transformations. With these definitions we now consider the functionals
\begin{align}
\label{Adef}
\A=&4r Z_X+\phi' [r^2 G_{3X}+G_{5X}(1-3f)-2 X f G_{5XX}+12 f  X (5 F_5+2 X F_{5X})]\nonumber\\[2mm]
=&4r Z_X+ \phi'(r^2 G_{3X}+G_{5X})+2\sqrt{f} Y_X\,,\\[3mm]
\B=&r Z- f \phi' X G_{5X}+12 f \phi' X^2 F_5\nonumber\\
=&rZ+\sqrt{f}Y\,, \label{Bdef}
\end{align}
where we have used the no ghost condition (\ref{BH}) relating $F_4$ and $F_5$ for any beyond Horndeski theory.
With these definitions the field equations can be shown to take a surprisingly simplified form,
\begin{align} 
&X' \A = 2\left( \frac{h'}{h}- \frac{f'}{f}\right)\B\,,\lab{eq1a}\\
&\frac{h' f}{2 h} \A= G_{2 X} r^2 + 2 G_{4X} -2 r f \phi' G_{3X}-2 f Z_X\,, \label{scalareqb1}\\
&2 f \frac{h'}{h} \B=- G_2 r^2 -2 G_4 -2 f Z\,.\lab{eq2a}
\end{align}

We can now make several general remarks on the form of the field equations that will be crucial when we attack the general case. Indeed if our theory enjoys parity symmetry ($G_3=G_5=F_5=0$) then the system is largely manageable, as we will explain in the next section, as $\A$ and $\B$ are related and \eqref{eq1a} is integrated directly. This is not the case once $G_3, G_5$ or $F_5$ are non zero and $\A$ and $\B$ are now independent. 

In the general case when \eqref{eq1a} is not integrable, we start by noting that the simplifying assumption $f=h$ already constrains heavily the theory at hand. Indeed, Eq. (\ref{eq1a}) imposes either $X$ to be constant or $\A=0$. 

When $X$ is constant, then (\ref{scalareqb1}) and (\ref{eq2a}) must give an identical solution for $h$. As we explicitly show in Appendix A, parity breaking theories render this impossible whereas even the parity-symmetric theories fail to lead to a viable solution. In fact, $X$ being a constant is a hypothesis trimmed to time (and radially) dependant scalar fields yielding quite often stealth solutions (see the nice general analysis in \cite{Kobayashi:2014eva}). We will therefore not consider any further the case of constant $X$.

We are therefore left with the latter condition, $\A=0$, which gives that the right hand side of (\ref{scalareqb1}) has to be  zero;
which is of course rather restrictive on the theory. 
If however this condition can be fulfilled, then the system is integrated by solving (\ref{eq2a}). 
We see therefore that the analysis is far more complex in parity asymmetric theories. 

There is however an example analytic solution of Horndeski theory \cite{Lu:2020iav} that follows the path we underlined above: For \cite{Lu:2020iav} we have the Horndeski theory functions $$G_2=8\alpha X^2,\; G_3=-8\alpha X,\; G_4=1+4\alpha X,\; G_5=-4\alpha \ln X$$ and for this case it turns out that the RHS of \eqref{psg} is simply
\begin{equation}
\label{psg}
 G_{2 X} r^2 + 2 G_{4X} -2 r f \phi' G_{3X}-2 f Z_X=\frac{\A}{f \phi'}=0  
\end{equation}
 giving that  the system of equations is actually compatible. We will come back and use this observation when studying parity breaking theories in section 4.

We will see throughout the following sections how to use equations (\ref{eq1a})-(\ref{eq2a}) to solve different cases.
It is useful to note that the degeneracy condition (\ref{BH}) relates $Z$ and $Y$ according to,
\begin{equation}
\label{ZY}
    Y=-\frac{Z}{2}\frac{G_{5X}(-2X)^{3/2}}{G_4-2X G_{4X}}.
\end{equation}

\setcounter{equation}{0}
\section{The integrability of parity symmetric theories}

In this section, we will consider shift symmetric beyond Horndeski theories with parity symmetry in $\phi$. We therefore take $G_3=G_5=F_5=0$. Several analytic solutions have been found for parity preserving (beyond) Horndeski theories mostly in the presence of a linear time-dependence  \cite{Babichev:2013cya}, \cite{Kobayashi:2014eva} , \cite{Charmousis:2014zaa}, \cite{Charmousis:2015aya}, \cite{Babichev:2015rva},\cite{Babichev:2016kdt} but also for a static scalar field \cite{Rinaldi:2012vy}, \cite{Anabalon:2013oea}, \cite{Minamitsuji:2013ura}, \cite{Cisterna:2014nua}, \cite{Babichev:2017guv}. Here, we will consider the resolution of such a theory with a static scalar field in full generality, and provide a method which will allow us to find concrete solutions. We will then pick out and look at certain interesting cases as illustrative examples.

In the absence of odd parity terms and with our auxiliary variables at hand, things greatly simplify since now $\A=4r Z_X$ and $\B=rZ$. It is then straightforward to integrate the first field equation (\ref{eq1a}),
to get
\begin{equation}
\label{rel1}
 Z^2 f= \gamma^2 h\,,   
\end{equation}
where $\gamma$ is a real constant that sets the relation  between $f$ and $h$. We remind the reader that $Z$ gives the relation between $F_4$ and $G_4$ \eqref{zdef}. The remaining field equations combine to give us
\begin{equation}
\label{rel2}
r^2 (Z G_2)_X+2 (G_4 Z)_X=0    
\end{equation}
and 
\begin{equation}
\label{rel3}
2 \gamma^2 (rh)'+Z(G_2 r^2+ 2 G_4)=0\,.  
\end{equation}
The former equation is essentially a condition on the theory which, once the theory is fixed, gives an algebraic relation determining $X$ (or equivalently the scalar field $\phi$). Note in passing that $X$ is only function of $r$ as the metric does not appear in this equation.  The last equation \eqref{rel3} is a first-order ODE for the metric function $h=h(r)$. Again, one has to fix the theory, i.e. choose the form of the coupling functions $G_2$, $G_4$ and $F_4$ (or $Z$), before proceeding to solve for $h$. The remaining metric function $f=f(r)$ may then be determined through (\ref{rel1}).
For later use, we note that the functional form of $Z$, i.e. choosing $Z=\gamma$ or $Z=Z(X)$, allows for the emergence or not, respectively, of homogeneous solutions with $f=h$. 

A general way to proceed in order to find explicit solutions is to consider an arbitrary function $\G=\G(X)$, such that 
\begin{equation}
\label{anz}
\G_{X}=\frac{\alpha r^2+ \beta}{\epsilon r^2+\delta} 
\end{equation}
and the field equation (\ref{rel2}) are compatible\footnote{Note that $\alpha$, $\beta$, $\epsilon$ and $\delta$ can in general be functions of $X$ but for simplicity we take them here as constants.}. Compatibility immediately gives the conditions
\begin{eqnarray}
G_2 Z= \epsilon \G- \alpha X + C\;,\quad
2 G_4 Z= \delta \G -\beta X + D\,, \label{rel22}
\end{eqnarray}
where $\epsilon$, $\delta$, $\alpha$, $\beta$, $C$ and $D$ are constants. The latter equation (\ref{rel3}) reduces to the form
\begin{equation}
\label{rel33}
    2\gamma^2 (rh)'+ r^2(\epsilon \G- \alpha X + C)+\delta \G -\beta X + D=0\,.
\end{equation}
In this form, \eqref{anz} and \eqref{rel33} are the field equations conveniently written using $\G$ while disentangling $r$ and $X$. Once we choose the function $\G =\G(X)$, \eqref{anz} determines $X$, and thus $\phi'$, in terms of $r$. The same equation provides also $\G$ in terms of $r$, and then \eqref{rel33} is an explicit $r$-dependent ODE giving the solution for $h$. The chosen form of $\G$ determines through \eqref{rel22} also the theory, namely the forms of the coupling functions $G_2$ and $G_4$ modulo $Z$. The latter quantity is fixed relative to the desired relation between $f$ and $h$, namely $f=h$ or $f \neq h$. Choosing $Z$ appropriately, we find $f$ through \eqref{rel1} and fix the remaining coupling function $F_4$ thus obtaining a full solution to the given theory. 

There are numerous examples one can consider with relative ease, essentially depending on whether \eqref{rel33} is integrable or not. In what follows, we will focus on two particular cases, one with $Z=\gamma$ leading to a homogeneous solution with $f=h$ and one with $Z=Z(X)$ for which a non-homogeneous solution emerges.


\subsection{Black holes with a canonical Kinetic term}

We will first consider the case of a constant $Z$ and, in particular, set $Z=\gamma$. Then, from \eqref{rel1}, we immediately get $f=h$. Let us also choose a linear $\G_X=2\mu X+\zeta$. Integrating the latter with respect to $X$ and substituting in \eqref{rel22}, we obtain the functional forms of $G_2$ and $G_4$, namely
\begin{eqnarray}
G_2 &=& \frac{\epsilon \mu}{\gamma} X^2+\frac{\epsilon \zeta-\alpha}{\gamma} X - 2\Lambda\,,  \nonumber \\
G_4 &=& \frac{\delta\mu}{2\gamma} X^2 +\frac{\delta\zeta-\beta}{2\gamma} X + 1\,.
\end{eqnarray}
In the above, we have without loss of generality fixed the constant term in the expression of $G_2$ to a vacuum cosmological constant $\Lambda$. We have likewise fixed the corresponding term in the expression of $G_4$ to unity in order to restore the Einstein-Hilbert term. Note that $G_2$ contains a linear term in $X$, therefore the theory includes a canonical kinetic term for the scalar field.

Since $Z$ determines the relation between $F_4$ and $G_4$ via \eqref{zdef}, making a choice for  $Z$ imposes certain constraints on the form of these two coupling functions. For example, in Horndeski theory where $F_4=0$, solutions with $Z=\gamma$, or equivalently with $f=h$, are only possible for $G_4=-\gamma +\sqrt{-2X}$. In beyond Horndeski theory, fixing $Z=\gamma$ and employing $G_4$ given above, completely determines the form of $F_4$ as
\begin{equation}
F_4=\frac{\gamma+1}{4X^2}+\frac{\beta-\delta \zeta}{8\gamma X}-\frac{3\delta \mu}{8\gamma}\,.
\end{equation}

Turning now to the derivation of the solution, we first notice that \eqref{anz} can be easily solved for $X$ giving the result
\begin{equation}
\label{X_hom}
    X(r)=\frac{(\alpha-\epsilon \zeta)r^2+\beta-\delta \zeta}{2\mu (\epsilon r^2+\delta)}\,.
\end{equation}
Substituting $X$ and $\G$ into (\ref{rel3}), we obtain, after a tedious but straightforward integration, the solution
\begin{equation}
    h(r)=C_1 + C_2 r^2+C_3\frac{\arctan(\sqrt{\frac{\epsilon}{\delta}}r)}{ \sqrt{\frac{\epsilon}{\delta}}r}-\frac{2M}{r},
    \label{h_first0}
\end{equation}
where
\begin{equation*}
C_1=-\frac{1}{\gamma}+\frac{(\epsilon\zeta-\alpha)[\delta(\epsilon\zeta+\alpha)-2\beta \epsilon]}{8\gamma^2\epsilon^2 \mu}\;, \qquad C_3=\frac{(\beta \epsilon -\delta \alpha)^2}{8\gamma^2\epsilon^2 \mu \delta}\,,
\end{equation*}
and $M$ is an integration constant{\footnote{It is interesting to note that a Newtonian fall-off arises for the mass term in parity preserving theories in agreement with \cite{Babichev:2020qpr} which arrives at this conclusion using a generalised Kerr-Schild method.}}. Also, 
$$C_2=\frac{\Lambda}{3\gamma} +\frac{(\epsilon \zeta-\alpha)^2}{24\mu \epsilon \gamma^2}$$ 
stands for the effective cosmological constant which is corrected by the linear term in $G_2$. If we wish this quantity to vanish without having to fine-tune the parameters of the theory, we have to set independently $\alpha-\epsilon \zeta=\Lambda=0$. We then find $C_1=-\frac{1}{\gamma}$ and we take $\gamma=-1$ to attain asymptotic flatness. Note that in the Horndeski theories considered in \cite{Rinaldi:2012vy} and \cite{Anabalon:2013oea}, solutions similar to (\ref{h_first0}) were obtained, but with an always non-zero effective cosmological constant. Going to beyond Horndeski theories seems to be the key point in order to attain asymptotic flatness with such a profile.

We therefore focus on this asymptotically flat black hole metric with $f=h$, which reads
\begin{equation}
\label{solh-hom}
    h(r)=1+\frac{(\beta-\delta \zeta)^2 }{8\delta \mu} \frac{\arctan(\sqrt{\frac{\epsilon}{\delta}}r)}{\sqrt{\frac{\epsilon}{\delta}}r}-\frac{2M}{r}\,,
\end{equation}
with $\epsilon\delta>0$, while the scalar field satisfies the equation
\begin{equation}
\label{phi_hom}
    \phi'^2(r)=-\frac{(\beta-\delta \zeta)}{\mu (\epsilon r^2+\delta)}\,\frac{1}{h(r)}\,.
\end{equation}
Thus, the scalar field is real for $r$ spacelike, provided the quantities $\delta\mu$ and $\left(\beta-\delta\zeta\right)$ have opposite signs. The scalar field diverges at the point where $h(r)$ vanishes, however, the fundamental scalar quantity of the theory $X \equiv - h \phi'^2/2$ remains everywhere finite. Note also that $\phi$ becomes trivial at asymptotic infinity as expected. Taking the same limit of \eqref{solh-hom}, we obtain a Reissner-Nordstrom type of solution. The ADM mass is given by $M$ along with the vacuum contribution, namely
\begin{equation*}
    M_{tot}=M-\frac{\pi(\beta-\delta \zeta)^2 }{32\delta \mu}\sqrt{\frac{\delta}{\epsilon}}\,.
\end{equation*}
We observe that, even in the case where $M=0$, we do not obtain a vacuum spacetime due to the contribution from the $\arctan$ term (see also \cite{Rinaldi:2012vy} and \cite{Anabalon:2013oea}). In fact, we see that we still have a black-hole horizon if $-(\beta-\delta \zeta)^2/8<\delta\mu<0$. Otherwise, we have a naked singularity since, at $r=0$, we have $h(0)\neq 1$ with a vacuum mass term due to the non trivial scalar field.
The asymptotic solution at large distance is completed by a tidal charge term with
\begin{equation}
    Q^2 = -\frac{(\beta - \delta \zeta)^2}{8 \mu \epsilon}\,.
\end{equation}
 For $M\neq 0$, there are straightforward constraints on its sign and magnitude so as to ensure a positive $M_{tot}$. Note in particular that $\delta \mu>0$ only allows for positive values of $M$, while negative values are also permitted if $\delta\mu<0$. Then, for $M>0$, there is a unique horizon and the spacetime is a black hole, while for $M<0$, the spacetime describes either a black hole with an outer and an inner horizon, or a naked singularity, depending on the parameters of the theory and on the magnitude of $M$ compared to the latter. 
Note that for $\delta\mu<0$, our solution possesses a robust Reissner-Nordstrom asymptotic limit with $Q^2>0$. 

With the above conditions, the solution \eqref{solh-hom}-\eqref{phi_hom} is therefore an asymptotically flat black hole with secondary scalar hair. As mentioned earlier, the $G_2$ term features in its expression a canonical kinetic term{\footnote{ Note that even in the absence of a linear term in $G_2$, substituting $X^2\longrightarrow -X$ restores the canonical kinetic term without changing the form of the equations and the metric solution itself. This is due to the form of the equations (\ref{rel2}-\ref{rel3}) and is not generally valid beyond parity symmetric theories}}. An interesting question arises in the context of the above theory regarding black holes and no-hair theorems: under what conditions on the theory can one include a canonical scalar kinetic term while having a non trivial black hole? According to  \cite{Babichev:2017guv}, in the case of spherical symmetry this is indeed possible if $G_4$ has a $\sqrt{-X}$ term included in its expression. Such a term was shown to provide a source term in the scalar equation in order for $\phi'$ to be non trivial \cite{Babichev:2017guv}, and to have a similar effect to the one of the $\phi$-Gauss-Bonnet term discussed in detail in \cite{Sotiriou:2014pfa}.
Our analysis here has demonstrated that hairy black holes can emerge for alternative forms of the coupling function $G_4$ while $G_2$ continues to feature a canonical kinetic term. This is due to an overall freedom regarding the choice of the form of the function $\G(X)$, which in turn defines the forms of $G_2$ and $G_4$. Both the solution presented here and the BCL one \cite{Babichev:2017guv} belong to the same class of hairy black holes which emerge in the context of beyond Horndeski theories described by the relations \eqref{rel22} but with different choices\,\footnote{We can limit ourselves to the Horndeski case with $F_4=0$ and obtain the BCL solution \cite{Babichev:2017guv} upon choosing $\G=\mu X+\zeta \sqrt{-2X}+\eta$ and fixing accordingly the constants of the theory.} for $\G(X)$. In support of this, in Appendix B, we present additional examples of hairy black holes with a Reissner-Nordstrom asymptotic behaviour (but with tidal rather than electric charge); these solutions emerge in beyond Horndeski theories with different forms of $G_4$ and $G_2$ functions but with always a canonical kinetic term for the scalar field.


Before closing this section, let us take a brief look at the solutions above in the presence of a positive or negative cosmological constant. In this case, $C_1$ need not be fixed to unity as the $C_2 r^2$-term is always dominant for large enough $r$. We can rather choose to fix $C_1+C_3=1$ so as, for $M=0$, the solution is regular at $r=0$, $h(0)=1+{\cal O}(r^2)$. By fixing the constants in such a way the solution close to $r=0$ has a (anti)de-Sitter core and we have no longer a solid deficit angle. The vacuum solution is therefore a regular but not maximally symmetric solution (de Sitter or anti de Sitter) because of the presence of the scalar field which continues to be non-trivial. In this case, we have therefore a soliton solution, i.e. an everywhere regular solution (such solutions have been found in higher order Proca theories \cite{Babichev:2017rti}). Adding a non trivial mass term gives us a black hole of zero electric or magnetic charge but with similar spacetime properties as the (A)dS-Reisser-Nordstrom black hole.  

\subsection{Non-Homogeneous Black holes}

In this case, we will assume that $Z=Z(X)$ and as a result obtain, via \eqref{rel1}, the relation
\be
f(r) =\frac{\gamma^2 h(r)}{Z^2(X)}\,.
\ee
The line-element then reads
\be
ds^2=-h(r) dt^2 + \frac{ Z(X)^2 dr^2}{\gamma^2 h(r)} +r^2 d\Omega^2\,.
\ee
There are numerous possibilities for the choice of $Z=Z(X)$ which in fact give different types of black hole, wormhole  or singular solutions. For a start if $Z=Z(X)$ is infinite at a particular finite radius $r=r_T$ (which is not a zero of $h$ and $f$) then $r=r_T$ is a possible wormhole throat (see \cite{Bakopoulos:2021liw}). But it is now easy to see that since $X$ is only function of $r$, according to \eqref{rel1}, $Z$ can only be a function of $r$ and crucially not of the metric functions $f$ and $h$. As such any throat will solely depend on the theory and not on an independent integration constant (essentially the mass parameter) which renders such wormholes eternal and fine tuned to the theory in question. We will not consider such a possibility furthermore. On the other hand, by choosing a smooth $Z(X)$ functional without zeros or singularities, we can obtain a non-homogeneous black-hole geometry (depending on the properties of $f$ and $h$). 

Here, we will focus on the latter case and, for simplicity, assume again  that  $\G_X=2\mu X+\zeta$. This means that our ODE for $h$, \eqref{rel33}, is identical to the one found in the previous subsection giving again the functional form \eqref{h_first0} with the same ($C_1, C_2, C_3$) theory-dependent constants. Similarly, \eqref{anz} gives us $X$ as the same function of $r$, \eqref{X_hom}, as before. We now choose an analytic and non-trivial $Z$ such that
\begin{equation}
\label{Z-nonhom}
    Z(X)=\gamma(1+X).
\end{equation}
Note that we have chosen $Z$ specifically aiming to obtain asymptotically flat black-hole solutions. Indeed, the solution for $h(r)$, \eqref{h_first0}, reduces to the asymptotically-flat Reissner-Nordstrom background, \eqref{solh-hom}, under the same choices made in the previous subsection, namely $\alpha - \zeta \epsilon=0$ and $\gamma=-1$. In that case, from \eqref{X_hom} we obtain
$$X(r)=\frac{\beta-\delta \zeta}{2\mu(\epsilon r^2+\delta)}.$$
Employing the above, the solution for $h(r)$, \eqref{solh-hom}, is completed by the expressions for $\phi'(r)$ and $f(r)$ given by
\begin{equation}
\phi'^2(r)=-\frac{(\beta-\delta \zeta)}{\mu (\epsilon r^2+\delta)}\,\frac{1}{f(r)}\,, \qquad f(r)=\frac{h(r)}{(1+X)^2}\,.
\end{equation}
We observe that $X$ goes to zero for large $r$ so that also $f(r)$ reduces asymptotically to unity, a result which validates our choice \eqref{Z-nonhom} from $Z(X)$. Since the solution for $h(r)$ is the same as in the previous subsection, the spacetime geometry preserves all the characteristics discussed there. Apart from the asymptotic flatness,  the spacetime features a tidal charge and one or two horizons or a naked singularity, following the description given in the previous subsection. 

The choice \eqref{Z-nonhom} for the functional $Z(X)$ fixes also the forms of the coupling functions $G_2$, $G_4$ and $F_4$, through \eqref{rel22}, as follows
\begin{align}
G_2  &=\frac{\epsilon \mu X^2}{\gamma (1+X)}\,, \nonumber \\[2mm]
 G_4  &= \frac{\delta \mu X^2+(\delta \zeta -\beta) X-2\gamma^2}{2\gamma (1+X)}\,,\nonumber\\[2mm]
 F_4 &= \frac{\beta -\delta  \zeta +X^2 \left(2 \gamma ^2-\delta  \mu \right)+X
   \left(-\beta +6 \gamma ^2+\delta  (\zeta -3 \mu )\right)}{8 \gamma  X (X+1)^2}\,.\nonumber
\end{align}
In the above expressions, we have set, for simplicity, $C+\epsilon \eta=0$ and $D+\delta \eta=-2\gamma^2$. At large distances, the above choices justify the vanishing of the cosmological constant and ensure the restoration of the Einstein-Hilbert term upon setting also $\gamma=-1$.

Having completed the presentation of black-hole solutions in the context of the parity-symmetric beyond Horndeski theory and before moving to the non parity-symmetric sector in the next section, we would like to stress that solutions with an asymptotic (A)dS-Reissner-Nordstrom behaviour arise also in cases where a mixed selection of coupling functions is made. In appendix C, we study a particular case where we keep $G_3=G_5=0$ but nevertheless allow for a non-vanishing $F_5$ term together with the $G_2$ and $G_4$ functions.


\setcounter{equation}{0}
\section{Attacking the case of no parity symmetry}

When we give up parity symmetry, keeping $G_3$, $G_5$ and eventually $F_5$ terms in the theory along with $G_2$, $G_4$ (and eventually $F_4$), the field equations get considerably more difficult to tackle in all generality. In the sense encountered in the previous section, integrability is lost and most known black hole solutions have been found numerically. To our knowledge, the only analytical solution is that of \cite{Lu:2020iav} valid in a parity breaking {\it{but shift symmetric theory}}. In order to extend the existing analytic results, we will try to generalize the theory appearing in \cite{Lu:2020iav}, where the solution obeys the condition $f=h$. As we commented also in section 2, the field equation (\ref{eq1a}) dictates that the choice $f=h$ is possible only if $\A=0$ or $X'=0$. The latter case was dealt with in appendix A and shown not to lead to a viable solution. When is then the former choice, $\A=0$, compatible with the remaining 2 field equations, or, to set the question in a different way, which theories allow for solutions with $f=h$? The answer is quite simple given the form of our field equations \eqref{eq1a}-\eqref{eq2a}.

Indeed, if we assume that $f=h$ and thus $\A=0$, then the RHS of (\ref{scalareqb1}) must also vanish. In this case, we may impose the condition
 \begin{equation}
 \label{rhs_cond}
      G_{2 X} r^2 +2G_{4X}-2 r f \phi' G_{3X}-2f Z_X=-\sqrt{f}\A Q,
 \end{equation}
where $Q=Q(X)$ is an arbitrary function of $X$, in order for the system to be well defined. Indeed $\A=0$ will be giving us the scalar field or function $X$ algebraically similarly to \eqref{rel2} and will be solving 2 out of the 3 field equations. The above equation can be seen as a polynomial in powers of $r$ (and $f$) with $X$-dependent coefficients. Employing the definition \eqref{Adef} for $\A$, and matching the corresponding coefficients of $r$ and $f$, \eqref{rhs_cond} leads to the  following universal constraints that must be valid for the theory functions
\begin{align}
  &  G_{2 X}=-\sqrt{-2X} Q G_{3 X}=-2Q^2 Z_X, \label{g2e}\\[3mm]
&  2 G_{4 X}=-\sqrt{-2X} Q G_{5 X}\label{g4e},\\[3mm]
&  Z_X=Q Y_X\,. \label{peq}
\end{align}
Combining the relation \eqref{g4e} with the no-ghost constraint (\ref{ZY}) gives
\begin{equation}
\label{Con}
    Z=Q Y\left(1- \frac{G_4}{2X G_{4X}}\right).
\end{equation}
Then, the compatibility of (\ref{peq}) with (\ref{Con}) leads to the additional constraint
\begin{equation}
\label{constraint}
    Q_X Y=\left(QY \frac{G_4}{2 X G_{4X}} \right)_{,X}\,.
\end{equation}
Let us then summarise: for the class of theories where the condition \eqref{rhs_cond} holds, the coupling functions $G_2$, $G_3$ and $G_5$ are given in terms of $G_4$, $Q$ and $Y$ via\footnote{Instead of $Y$, which involves the odd coupling functions $G_5$ and $F_5$, one may equivalently choose $Z$ as the third variable, which features the even functions $G_4$ and $F_4$.} (\ref{g2e})-(\ref{g4e}). Due to the additional constraint \eqref{constraint}, only two of the latter three quantities are independent and thus our solutions will be parametrised by two free theory functions, say $G_4$ and $Q$. 

We may therefore start our analysis by choosing $Q=Q(X)$ and $G_4$, determining $Y$ via \eqref{constraint}. Then, (\ref{g2e})-(\ref{peq}) will completely fix the beyond Horndeski theory by providing the forms of the coupling functions $G_2$, $G_3$, $G_5$, $F_4$ (via $Z$) and $F_5$ (via $Y$). This theory choosing filter is clearly not the most general parity breaking theory but it is clearly a general starting point admitting \cite{Lu:2020iav} as one particular solution as we will see. 

The field equations \eqref{eq1a}-\eqref{eq2a}  for this special class of theories simplify to:
\begin{align} 
&X' \A = 2\left(\frac{h'}{h} - \frac{f'}{f}\right)\B\,,\lab{eq1aa}\\
&\A \left(\frac{h' \sqrt{f}}{h}+2Q \right)=0\,, \label{scalareqb2}\\
&2f \frac{h'}{h} \B +G_2 r^2 +2 G_4 +2 f Z=0\,,\lab{eq2a2}
\end{align}
where \eqref{scalareqb1} was rewritten with the help of \eqref{rhs_cond}. If we look for solutions with $f=h$, then \eqref{eq1aa} leads immediately to $\A=0$; $\A$ can be conveniently rewritten, by combining its definition \eqref{Adef} with the constraints (\ref{g2e})-(\ref{g4e}), as
\begin{equation}
    \A=\frac{2 Z_X}{Q\sqrt{f}}\left[(r Q+\sqrt{f})^2- \frac{G_{4X}}{Z_X} \right]. \label{eqascalarfieldcompatible}
\end{equation}
The vanishing of the above combination will then give us the solution for $X$ (or the scalar field $\phi'$) as a function of $f$. Note here a crucial difference with parity symmetric theories where we saw that $X$ only depended on $r$ \eqref{rel2} and not on the metric functions $f$ or $h$. Equation \eqref{scalareqb2} will be trivially satisfied, for $f=h$, whereas  \eqref{eq2a2} will provide a first-order differential equation for the sole metric function $f(r)$. If one solves this latter ode a full solution will be known to this theory. 

Let us note that, although the condition \eqref{rhs_cond} was motivated by the assumption that $f=h$, the field equations \eqref{eq1aa}-\eqref{eq2a2} allow also for the emergence of solutions with $f\neq h$ within the same class of theories; in that case, $\A \neq 0$ and the solution for the scalar field follows instead from the equation $\frac{h' \sqrt{f}}{h}+2Q =0$ -- provided that $Q(X)$ is not trivial -- leaving us with two ODEs, \eqref{eq1aa} and \eqref{eq2a2}, to determine the two unknown metric functions $h$ and $f$.

In the context of the present analysis, we will focus on homogeneous solutions with $f=h$. As outlined above, the first step towards finding such a solution is to choose the form of $Q(X)$. To this end, we set $Q(X)=\gamma \,(-2X)^m$, with $\gamma$ a constant of dimension $2m-1$. This choice, in conjunction with \eqref{constraint} and \eqref{peq}, entails the general solution for all $m$,
\begin{equation}
    Y(X)=c (-2X)^{1-m}\,G_{4X} \,G_4^{2m-1}, \qquad Z(X)=c\gamma \,G_4^{2m-1}(G_4-2 X G_{4X})\,,
\end{equation}
where $c$ is a constant of integration and the product $\gamma c$ is dimensionless. 
The above relations determine the functionals $Y$ and $Z$ in terms of $G_4$. Fixing the latter as well as the value of $m$ will allow us to completely fix the theory via \eqref{g2e}-\eqref{g4e} and then integrate \eqref{eq1aa}-\eqref{eq2a2} in order to find the solution. For this purpose, in the next two subsections we choose to study two separate cases: the case with $m=1/2$ leading to $Q=\gamma\sqrt{-2X}$ and the case with $m=0$ corresponding to $Q=\gamma$ an arbitrary constant. The first case is particularly interesting as it corresponds to (Horndeski) theories related to the Kaluza Klein reduction of Lovelock theories \cite{VanAcoleyen:2011mj}, \cite{Charmousis:2012dw}, and includes the shift symmetric theory and black hole found in \cite{Lu:2020iav}. 


\subsection{Parity breaking theories related to Kaluza-Klein reduction of Lovelock theory}\label{first_q}

If we choose $m=1/2$, then we obtain  $Q=\gamma\sqrt{-2X}$, and $\gamma$ is a dimensionless quantity. Subsequently, one finds:
$$
Y=c\; G_{4X} \sqrt{-2X}, \qquad Z=\gamma c\,(G_4-2 X G_{4X})\,.
$$
Without further ado we may now completely fix the theory by choosing $G_4 = 1+\alpha\left(-2X\right)^n$. 
In fact the case $n=1, \gamma c=-1$ gives the theory \cite{Lu:2020iav} that originates from higher dimensional Lovelock theory and in particular the Kaluza Klein reduction of the Gauss-Bonnet term{\footnote{See also \cite{Fernandes:2021dsb} for an alternative interesting derivation}}. One can loosely relate $n+1$ to the order, in powers of curvature, of the higher order Lovelock term{\footnote{Although a complete justification of this is beyond the scope of this paper it is nevertheless an important characterisation of $n$}} involved in the higher dimensional reduction. For example, the $n=2$ theory can be seen to be related to the third order Lovelock term $L_3$ \cite{Alkac:2022fuc} via Kaluza klein reduction to 4 dimensions. 
In particular, for $n=2$, we have $G_4=1+\alpha X^2$ and $G_5=X$ which interestingly gives the self tuning Paul term \cite{Kobayashi:2019hrl} found in Fab 4 \cite{Charmousis:2011bf, Charmousis:2011ea}.

The remaining coupling functions of the theory are then:
\begin{align*}
    G_2 ={}& 2\gamma^3c\alpha n (2n-1)\frac{\left(-2X\right)^{n+1}}{n+1},\\[1mm] G_3 ={}& -2\gamma^2c\alpha (2n-1)\left(-2X\right)^n,\quad G_{5X}=\frac{4\alpha n}{\gamma}\left(-2X\right)^{n-2}\,, \\ 
    F_4 ={}& \frac{\gamma c+1}{4X^2}\left(1+\alpha(1-2n)\left(-2X\right)^n\right),\quad F_5 = \frac{\gamma c+1}{3\gamma}\left(-2\alpha n\right)\left(-2X\right)^{n-3}\,.
\end{align*}
As before, we have chosen not to include a cosmological constant in $G_2$. The Horndeski case is readily identified with the choice $\gamma c=-1$ for which both $F_4$ and $F_5$ vanish. In addition, the choices $n=1$ and $\gamma=1$ correspond precisely to the case of  \cite{Lu:2020iav} (for our conventions on the coupling constants).
Here, we will keep $n$ and $\gamma c$ arbitrary and attempt to generalise, in the context of beyond Horndeski theory, the black-hole solution found in \cite{Lu:2020iav}. 

The constraint $\mathcal{A}=0$, valid for all solutions with $f=h$, in conjunction with  (\ref{eqascalarfieldcompatible}) lead to the following relation for the scalar field:
\begin{equation}
\label{phi_Q1}
    \phi'=\frac{1-\sqrt{\gamma c(1-2n)f}}{r\gamma\sqrt{\gamma c(1-2n)f}}\,.
\end{equation}
The above clearly demands that $n\neq 1/2$ and that we define a positive function $F^2(r) \equiv \gamma c(1-2n)f>0$. Then, the differential equation (\ref{eq2a2}) for the metric function can be integrated once with respect to the radial coordinate $r$ to give the following equation:
\begin{equation}
    (n+1)\left(\gamma^3c(1-2n)\right)^n r^{2n}\left(1-2n+F^2\right)+\alpha\left(1-F\right)^{2n}\left(1+2nF+F^2\right)-\lambda r^{2n-1}=0\,, \label{eqalgsqrt}
\end{equation}
where $\lambda$ is an integration constant. The above is an algebraic equation with $2(n+1)$ degree in $F$. For $n\geq 1$, integer or half-integer, it becomes a polynomial equation in $F$. 
In the case where $n=1$, we easily obtain the explicit result:
\begin{equation}
    f(r) = -\frac{1}{\gamma c}-\frac{r^2\gamma^2}{\alpha}\left(1\pm\sqrt{1+\frac{\alpha\lambda}{r^3\gamma^6c^2}}\right).
\end{equation}
The above, as anticipated from our earlier comments, reduces to the black-hole solution \cite{Lu:2020iav} when $\gamma c=-1$ (Horndeski case). Note that the extension of the solution in beyond Horndeski picks up a solid angle deficit since $\gamma c \neq-1$ while $\phi'\neq 0$ for $M=0$. In other words if $\gamma c \neq-1$ and $M=0$ we have a naked singularity at $r=0$ while the scalar field also explodes there. It is only the Horndeski case which has a well defined vacuum for $M=0$.  

Let us now move on to the general $n$ case. Equation (\ref{eqalgsqrt}) implies the following asymptotic behaviour for the metric function:
\begin{equation}
    f(r)=-\frac{1}{\gamma c}+\frac{\lambda}{(n+1)\gamma c (1-2n)\left(\gamma^3c(1-2n)\right)^n}\frac{1}{r} +\mathcal{O}\left(\frac{1}{r^{2n}}\right). \label{eqasymplupanggeneral}
\end{equation}
Employing the above into \eqref{phi_Q1}, we find in turn that the scalar field behaves at infinity as
\begin{equation}
    \phi'=\frac{1-\sqrt{2n-1}}{r\gamma\sqrt{2n-1}}+\mathcal{O}\left(\frac{1}{r^2}\right),
\end{equation}
which implies that $\phi$ diverges like $\ln(r)$ at infinity, except for $n=1$ where $\phi=\mathcal{O}\left(\frac{1}{r}\right)$. However, $-2X=f \phi'^2$ always vanishes at infinity. As regards the metric, it is asymptoting Minkowski spacetime only for the Horndeski case ($\gamma c=-1$){\footnote{If $\gamma c\neq -1$ the asymptotic metric is only locally asymptotically flat. By this we mean that the spacetime curvature tensor asymptotes zero but one may have a global deficit angle as in the case of the gravitating monopole solution in GR \cite{Barriola:1989hx}}}. In this case, we may restore a Schwarzschild-like behaviour\footnote{In fact, one can notice on (\ref{eqasymplupanggeneral}) that the metric is almost Schwarzschild at infinity for large values of $n$.} if the integration constant $\lambda$ is related to the mass $M$ by
\begin{equation}
    \lambda = 2M(n+1)(1-2n)\left(\gamma^2(2n-1)\right)^n.
\end{equation}
In what follows we will focus on the Horndeski case, and thus take $\gamma c=-1$, but allow for general $n$. We will also set $\gamma=1$, since $\gamma$ can always be absorbed into $\alpha$ via the rescaling  $\alpha\to \alpha/\gamma^{2n}$. All the above combined allow to write \eqref{eqalgsqrt} in the form:
\begin{equation}
     (n+1)\left(2n-1\right)^n r^{2n-1}\left[(2n-1)(2M-r)+rF^2\right]+\alpha\left(1-F\right)^{2n}\left(1+2nF+F^2\right)=0\,. \label{eqsimplen}
\end{equation}
Although we cannot solve analytically (\ref{eqsimplen}) for $n>1$, we can find a perturbative solution assuming $\alpha\rightarrow 0$. To this end, we use the expansion 
\begin{equation}
\label{f_pert}
    f(r)=1-\frac{2M}{r} +\sum_{i=1}^\infty \alpha^if_i(r).
\end{equation}
By replacing the above expression in (\ref{eqsimplen}) and solving order by order, we can determine the functions $f_i(r)$. For instance, the first-order correction is found to be
\begin{align}
    &f_1(r)=-\frac{2\, [1-k(r)]^{2 n} \left[n r (k(r)+1)-2 M
   n+M\right]}{r^{2n+1}\,(n+1)\,(2 n-1)^{n+1}}\,,
\end{align}
where
\begin{equation}
    k(r)=\sqrt{(2n-1)\left(1-\frac{2M}{r}\right)}.
\end{equation}
The above perturbative form \eqref{f_pert} for the metric function $f(r)$ is valid over the entire radial regime in the small-$\alpha$ limit.

The presence of a horizon is signified by the vanishing of the metric function $f(r)$. Thus, setting $F(r)=0$ in \eqref{eqsimplen} defines the horizon radius as the value $r_H$ which satisfies the equation:
\begin{equation}
    (n+1)\left(2n-1\right)^{n+1} r_H^{2n-1}\left(r_H-2M\right)=\alpha\,. \label{eqhorizonn}
\end{equation}
As expected, the coupling parameter $\alpha$ induces a deviation from the Schwarzschild radius $r_0=2M$ (for clarity and in order to compare with the GR limit, we restrict here to the case $M>0$). As above, one can therefore write a series expansion for the horizon radius, in the limit $\alpha \rightarrow 0$, of the form
$$r_H = 2M + \sum_{k=1}^{\infty}b_k\,\alpha^k\,. $$
The coefficients $b_k$ can be found again order by order, with the first three given by the expressions
\begin{equation}
    b_1 = \frac{1}{(n+1)(2M)^{2n-1}(2n-1)^{n+1}},\quad b_2 = \frac{1-2n}{2M}b_1^2,\quad b_3 = \frac{(2n-1)(3n-1)}{(2M)^2}b_1^3\,.
\end{equation}
One can check that the same results follow by demanding the vanishing of the perturbative solution \eqref{f_pert}.

The presence of the coupling parameter $\alpha$ in \eqref{eqhorizonn} not only does change the horizon radius $r_H$ compared to GR but it also determines the number of roots of that equation, and thus the topological structure of spacetime. Equation \eqref{eqhorizonn} has the simple form of a polynomial of order $2n$, and much can be said about its real, positive roots. To this end, we define the quantities:
\begin{equation}
    r_n=\left(1-\frac{1}{2n}\right)2M<2M,\qquad \alpha_M=-(n+1)\left(2n-1\right)^{3n} \left(\frac{M}{n}\right)^{2n}<0\,.
\end{equation}
Then, if $\alpha>0$, there is a unique horizon with  $r_H>2M$. If $\alpha_M<\alpha<0$, there are exactly two horizons with $r_{H-}<r_n<r_{H+}<2M$. One has that $r_{H+}\to \left(2M\right)^-$ when $n\to\infty$, and $r_{H-}\to 0^+$ when $\alpha\to 0^-$. If $\alpha<\alpha_M$, the spacetime has no horizons. One can note that, as $n\to\infty$, $\alpha_M\sim -M\left(2n\right)^{n+1}\left(2M\right)^{2n-1}\exp\left(-3/2\right)\to -\infty$, hence a larger and larger parameter space emerges for $\alpha$ which allows for black-hole solutions. All the previous results are consistently illustrated by the Lu-Pang case \cite{Lu:2020iav}, which has $n=1$, where:
\begin{equation}
    r_{H\pm} = M\pm\sqrt{M^2+\frac{\alpha}{2}}\,.
\end{equation}

It is worth noting that the case $\alpha<0$ always leads, for any $n$, to a positive $f$ at the origin: either with a black-hole metric function $f(r)$ changing sign twice if $\alpha>\alpha_M$, or with a soliton if $\alpha<\alpha_M$. The behaviour of the metric function as $r\to 0$ can be derived from  equation (\ref{eqsimplen}), and reads:
\begin{equation}
    f(r) = \frac{1}{2n-1}-\frac{2}{2n-1}\left(\frac{M}{-\alpha}(2n-1)^{n+1}\right)^{\frac{1}{2n}}r^{1-\frac{1}{2n}}+\mathcal{O}\left(r^{1-\frac{1}{2n}}\right). \label{eqrto0n}
\end{equation}
We observe that a solid angle deficit emerges, which is however covered by the horizons when $\alpha_M<\alpha<0$.


\subsection{Theories admitting solutions with non trivial asymptotics}\label{second_q}
We will now consider the case $m=0$ which leads to  $Q=\gamma$, with $\gamma$ a constant of dimension $-1$. This case is not particularly physically motivated as in the previous section, but has some simplifying mathematical properties. We now obtain:
\begin{equation}
    Y = c(-2X)\frac{G_{4X}}{G_4},\qquad Z = \gamma(c+Y)\,.
\end{equation}
In order to fix the theory, we will take again $G_4 = 1+\alpha\left(-2X\right)^n$. The remaining coupling functions of the theory are now:
\begin{align*}
    G_2 ={}&-2\Lambda-2\gamma^3c\frac{1+\alpha(1-2n)\left(-2X\right)^n}{1+\alpha\left(-2X\right)^n}\,,\quad G_{3X} = 8\gamma^2c\alpha n^2\frac{\left(-2X\right)^{n-3/2}}{\left(1+\alpha\left(-2X\right)^n\right)^2}\,, \\[1mm] G_{5X}={}&\frac{4\alpha n}{\gamma}\left(-2X\right)^{n-3/2}\,,\quad 
    F_4 =\frac{\left(1+\gamma c+\alpha\left(-2X\right)^n\right)\left(1+\alpha(1-2n)\left(-2X\right)^n\right)}{4X^2\left(1+\alpha\left(-2X\right)^n\right)},\\[1mm] F_5 ={}& \frac{-2n\alpha}{3\gamma}\left(-2X\right)^{n-5/2}\frac{1+\gamma c+\alpha\left(-2X\right)^n}{1+\alpha\left(-2X\right)^n}\,,
\end{align*}
where a cosmological constant $\Lambda$ is now included in $G_2$. In this case, we observe that, for $G_4=G_4(X)$, the functions $F_4$ and $F_5$ are always non-trivial, hence the $Q=\gamma$ case always belongs to beyond Horndeski theories\footnote{The opposite holds true only in the less interesting $n=0$ case, and thus for a trivial $G_4$. Then, for generic coefficients we obtain general relativity with an $F_4$ correction. The latter can be  eliminated upon making an appropriate choice of coefficients, namely $1+ \gamma c + \alpha=0$.}. Note also that any $n$ leads to algebraic functions of $X$, except $n=1/2$, which gives a logarithmic $G_5$ (i.e. a Gauss-Bonnet term) and a logarithmic $G_3$.

Since we assume again that $f=h$, \eqref{eq1aa} dictates that $\mathcal{A}=0$ and (\ref{eqascalarfieldcompatible}) gives:
\begin{equation}
    \left(r\gamma+\sqrt{f}\right)^2=\frac{G_4^2}{-2n\gamma c}\,, \label{A=0}
\end{equation}
which implies that $-2n\gamma c>0$. Therefore, we can express $G_4$ directly with respect to $r$ as $G_4 = \beta  \left(r\gamma+\sqrt{f}\right)$, with $\beta=\pm\sqrt{-2n\gamma c}$. With this information at hand (\ref{eq2a2}) can then be integrated once directly to give the following third order polynomial equation in $\sqrt{f}$:
\begin{equation}
    4ncf^{3/2}+3\left(\frac{\beta}{\gamma}+\left(2n-1\right)r\gamma c\right)f+r^2\left(r\Lambda_n-3\beta\gamma\right)-\lambda=0\,. \label{eq3rdorderf}
\end{equation}
In the above, $\lambda$ is an integration constant with dimension 1, and we have introduced the quantity $\Lambda_n=\Lambda+\gamma^3 c\left(1-2n\right)$ for simplicity. Interestingly, the coupling parameter $\alpha$ of $G_4$ does not play any role in \eqref{eq3rdorderf}. We notice that, due to the $f^{3/2}$ term, this equation gives by construction the form of $f$ only in the spacetime regions where $f\geq 0$. Let's look for solutions where $f$ is positive at infinity. The asymptotic behaviour is then an Anti-de Sitter one:
\begin{equation}
    f(r) = \Lambda_f r^2 + \epsilon_1 r+\epsilon_0+\frac{\epsilon_{-1}}{r}+\mathcal{O}\left(\frac{1}{r^2}\right) \label{eqasymp}
\end{equation}
where the coefficient $\Lambda_f>0$ satisfies the equation
\begin{equation}
    4nc\Lambda_f^{3/2}+3\left(2n-1\right)\gamma c\Lambda_f+\Lambda_n=0\,, \label{eqpolycosmo}
\end{equation}
and the coefficients $\epsilon_i$ are given by
\begin{align}
    \epsilon_1={}& \Xi  \left(\beta\gamma-\frac{\beta\Lambda_f}{\gamma}\right), \label{eqepsilon1}\\
    \epsilon_0={}&\Xi \left(-\frac{\beta\epsilon_1}{\gamma} -\frac{nc\epsilon_1^2}{2\sqrt{\Lambda_f}}\right),\label{eqepsilon0}\\
    \epsilon_{-1}={}&\Xi \left(-\frac{\beta\epsilon_0}{\gamma} -\frac{nc\epsilon_1\epsilon_0}{\sqrt{\Lambda_f}}+\frac{nc\epsilon_1^3}{12\Lambda_f^{3/2}}+\frac{\lambda}{3}\right).
\end{align}
In the above, we have defined the quantity
$\Xi = \left[2nc\sqrt{\Lambda_f}+\left(2n-1\right)\gamma c\right]^{-1}$, for simplicity. The equation (\ref{eqpolycosmo}) giving $\Lambda_f$ admits generic solutions depending on the exact values and relative signs of $nc$ and $\Lambda_n$. 

The presence of horizons with $r_H>0$ is linked again with the vanishing of $f$, in which case \eqref{eq3rdorderf} gives the constraint
\begin{equation}
    r_H^2\left(r_H\Lambda_n-3\beta\gamma\right)-\lambda=0\,.
\end{equation}
This third order polynomial can be easily analysed. We will consider the case $\beta\gamma<0$ as the  sign of this quantity does not affect the type of spacetime that emerges. Then, defining $\lambda_M = -4\left(\beta\gamma\right)^3/\Lambda_n^2$, we  find that:
\begin{itemize}
    \item If $0<\lambda<\lambda_M$ and $\Lambda_n>0$, there is exactly one horizon. Therefore the function $f$, which is positive at infinity, will change sign once and be negative at the origin.
    \item If $0<\lambda<\lambda_M$ and $\Lambda_n<0$, there will be exactly two horizons $r_{H-}<2\beta\gamma/\Lambda_n<r_{H+}$: $f$ will be positive both below $r_{H-}$ and above $r_{H+}$.
    \item If $\lambda<0$ or $\lambda>\lambda_M$, there will be a unique horizon if $\lambda\Lambda_n>0$ (thus a similar behaviour as for the first case), and no horizon otherwise.
\end{itemize}

It is worth investigating also the behaviour of the metric function as $r\to 0$, in the cases where $f$ is positive there. In that case, its form is given by:
\begin{equation}
    f(r)=\eta_0+\frac{\gamma^2c\left(1-2n\right)}{\beta+2n\gamma c\sqrt{\eta_0}}\,\eta_0 r + \mathcal{O}\left(r^2\right), \label{eqorigin}
\end{equation}
provided that $\eta_0$ satisfies
\begin{equation}
    4nc\eta_0^{3/2}+\frac{3\beta}{\gamma}\eta_0-\lambda=0\,. \label{eqeta0}
\end{equation}
The above equation admits solutions in various cases, for example by setting $nc\lambda>0$.
One sees that a de Sitter core arises for the case $n=1/2$ (i.e. the Gauss-Bonnet choice): demanding that $f(0)=\eta_0=1$, which fixes $\lambda=2c+3\beta/\gamma$ from (\ref{eqeta0}), the metric function as $r \rightarrow 0$ reads: 
\begin{equation}
    f(r)=1+\frac{\beta\gamma^2}{\beta+\gamma c}r^2-\frac{\Lambda\gamma}{3\left(\beta+\gamma c\right)}r^3+\mathcal{O}\left(r^4\right).
\end{equation}
Meanwhile, from \eqref{eqpolycosmo}, the cosmological constant at infinity takes the simple form:
\begin{equation}
    \Lambda_f = \left(-\frac{\Lambda}{2c}\right)^{2/3}, \label{eqcosmosimple}
\end{equation}
since in this case $\Lambda_n=\Lambda$.
If a black hole were to respect both these forms at $r\to 0$ and $r\to \infty$, it would belong to the second case described above. This is possible if $\gamma$, $c$ and $\Lambda$ exist such that $\beta \gamma<0$ (by symmetry), $\Lambda<0$ and $0<\lambda<\lambda_M$, which now reads: 
\begin{equation}
    0<2c+\frac{3\beta}{\gamma}<\frac{4\gamma^4 c\beta}{\Lambda^2}. \label{eqregular}
\end{equation}
For example, the choice $\gamma<0$, $\gamma c\equiv-\delta$ and $\beta=\sqrt{\delta}$, gives a middle term positive for any $\delta>9/4$, while the RHS is positive, and choosing a sufficiently small $\left\lvert\Lambda\right\rvert$ ensures the fulfilment of this constraint. The corresponding spacetime is a regular black hole with two horizons and an AdS behaviour at infinity.

Finally, let us discuss the profile of the scalar field, the first derivative of which follows from \eqref{A=0} and is given by:
\begin{equation}
    \phi'=\frac{1}{\sqrt{f}}\left(\frac{\beta\left(r\gamma+\sqrt{f}\right)-1}{\alpha}\right)^{\frac{1}{2n}}\underset{r\to\infty}{=}\left(\frac{\beta\left(\gamma+\sqrt{\Lambda_f}\right)}{\alpha}\right)^{\frac{1}{2n}}\frac{r^{\frac{1}{2n}-1}}{\sqrt{\Lambda_f}}+\mathcal{O}\left(r^{\frac{1}{2n}-2}\right)\,.
\end{equation}
Above, we have also derived the asymptotic form of $\phi'$ at radial infinity: it diverges for $0<n<1/2$, while it converges to a constant value if $n=1/2$, and to zero otherwise. However, whenever $\gamma=-\sqrt{\Lambda_f}$, the above behaviour is modified and becomes:
\begin{equation}
    \phi' = -\frac{\left(-1/\alpha\right)^{1/2n}}{r\gamma}+\mathcal{O}\left(\frac{1}{r^3}\right)
\end{equation} 
such that $-2X=f \phi'^2$ converges for any $n$. Given (\ref{eqpolycosmo}), the case $\gamma=-\sqrt{\Lambda_f}$ corresponds to a unique relation between the theory coefficients, namely $\Lambda=2\gamma^3 c$, with $\gamma <0$. This choice leads to a simplification of the asymptotic development (\ref{eqasymp}): $\epsilon_1$ and $\epsilon_0$ both vanish, consistently with (\ref{eqepsilon1}) and (\ref{eqepsilon0}), while $\Lambda_f$ recovers the simple expression (\ref{eqcosmosimple}) for any value of $n$. Finally, for $n=1/2$, a regular black hole can still be obtained, since the constraint (\ref{eqregular}) is verified for any $\delta\equiv-\gamma c\in\left]9/4,\delta_M\right[$, where $2\delta_M^{3/2}-3\delta_M-1=0$.

%

\setcounter{equation}{0}
\section{Disformal transformations and solution generation techniques}

It is well known that a disformal transformation $D$ depending on $X$ takes a solution of Horndeski theory to a solution of beyond Horndeski theory (see for example \cite{Crisostomi:2016tcp}, \cite{BenAchour:2016cay}, \cite{BenAchour:2016fzp}). In a recent publication \cite{Bakopoulos:2021liw}, such techniques were used to construct a traversable regular wormhole solution (see also \cite{Faraoni:2021gdl}, \cite{Chatzifotis:2021hpg}). Let us re-visit this construction here with the benefit of our simplified field equations \eqref{eq1a}-\eqref{eq2a}.  Following \cite{Bakopoulos:2021liw}, by barred quantities we will be noting the seed ``known" Horndeski solution. As such we have $\bar{\phi}, \bar{h}, \bar{f}$ and of course $\bar{X}=-\frac{1}{2}\bar{f}\bar{\phi'^2}$ for some specific set of coupling functions $\{\bar{G_i}\}$ in Horndeski theory. Via some $D(X)$ function we go to the ``unknown" image metric which is a solution in beyond Horndeski theory \{$G_i, F_4, F_5$\}, and reads
$$
g_{\mu \nu}=\bar{g}_{\mu \nu}-D(\bar{X})\, \partial_\mu \phi\, \partial_\nu \phi\,.
$$
Given that $\phi$ is only a function of $r$, we have immediately that $\bar{\phi}=\phi$, $\bar{h}=h$ whereas the only terms that do change for the image solution are the following
\begin{equation}
\label{disfbh}
f=\frac{\bar{f}}{1+2 D \bar{X}},\qquad X=\frac{\bar{X}}{1+2 D \bar{X}}\,.
\end{equation}

As we vary the disformability function $D$, we span all values of $f$ for the given same $h$ and $\phi$ while at the same time we change the theory according to specific  transformation rules. For the case of spherical symmetry, these rules read
\begin{eqnarray}
\label{disformal2}
G_4&=&\frac{\bar{G_4}}{(1+2 \bar{X} D)^{1/2}}\,,\qquad
G_{5X}=\frac{\bar{G}_{5\bar{X}}(1+2 \bar{X} D)^{5/2}}{1-2 \bar{X}^2 D_{\bar{X}}}\,,\label{g5d}\\
F_4&=&(\bar{G_4}-2\bar{X}\bar{G}_{4\bar{X}})\frac{D_{\bar{X}}(1+2 \bar{X} D)^{5/2}}{2(1-2 \bar{X}^2 D_{\bar{X}})}\,,\label{f4d}\\
F_5&=&\bar{X}\bar{G}_{5\bar{X}}\frac{D_{\bar{X}}(1+2 \bar{X} D)^{7/2}}{6(1-2 \bar{X}^2 D_{\bar{X}})}\,.\label{f5d}\\
G_2&=&\frac{\bar{G_2}}{(1+2 \bar{X} D)^{1/2}}\,,\qquad
G_{3X}=\bar{G}_{3\bar{X}}\frac{(1+2 \bar{X} D)^{5/2}}{1-2 \bar{X}^2 D_{\bar{X}}}\,.\label{g3d}
\end{eqnarray}
In the above, we have used the relation
\begin{equation}
    D_X=D_{\bar{X}} \frac{(1+2 \bar{X} D)^2}{1-2 \bar{X}^2 D_{\bar{X}}}\,,
\end{equation}
while the constraint (\ref{BH}) is verified.

Given our simplified general equations of motion \eqref{eq1a}-\eqref{eq2a}, one can easily find how our variables $Z, Y$, $\A$ and $\B$ transform under the disformal transformation. We find
\begin{eqnarray}
\label{disf2}
Z&=&(1+2 \bar{X} D)^{1/2}\bar{Z}\,,\quad Y=(1+2 \bar{X} D)^{1/2}\bar{Y}\,,\quad
\B=(1+2 \bar{X} D)^{1/2}\bar{\B}\,,\\[2mm]
\A&=&\frac{(1+2 \bar{X} D)^{5/2}}{1-2 \bar{X}^2 D_{\bar{X}}}\bar{\A}+4\frac{(1+2 \bar{X} D)^{3/2}}{1-2 \bar{X}^2 D_{\bar{X}}}(D+\bar{X}D_{\bar{X}})\,\bar{\B}\,. \label{disf3}
\end{eqnarray}
Hence, given a solution of $\{\bar{G}_i,\; i=2,3,4,5\}$-Horndeski theory with \{$\bar{f}, \bar{h},\bar{X}$\}, we obtain a solution of beyond Horndeski theory with $\{G_i, F_4, F_5,\; i=2,3,4,5\}$ given by  \eqref{disformal2}-\eqref{g3d} for any $D=D(X)$ with \{$f,X$\} given in \eqref{disfbh} and $h=\bar{h}$.
Using the above transformation rules \eqref{disf2}-\eqref{disf3}, it is straightforward to show that the field equations remain invariant under an arbitrary disformal transformation. The functional $D(X)$ determines the nature of the image solution. Let us examine some simple non trivial examples.

As it was shown in \cite{Bakopoulos:2021liw}, one can obtain a beyond-Horndeski wormhole solution starting from a Horndeski black-hole solution. The seed solution found by Lu and Pang  \cite{Lu:2020iav} (see also \cite{Hennigar:2020lsl}, \cite{Fernandes:2020nbq}) is described by the following Horndeski theory functions 
$$G_2=8\alpha \bar{X}^2,\; G_3=-8\alpha \bar{X},\; G_4=1+4\alpha \bar{X},\; G_5=-4\alpha \ln\bar{|X|}.$$
The solution reads \cite{Lu:2020iav}
\begin{equation}
   \bar{h}(r)=\bar{f}(r)=1+\frac{r^2}{2 \a}\left(1-\sqrt{1+\frac{8\a M}{r^3}}\right), \quad {\rm and} \quad \bar\f'=\frac{\sqrt{\bar{h}}- 1}{r \sqrt{\bar{h}}},\label{phih}
\end{equation}
and describes a black hole with ADM mass $M$ and a non trivial scalar field, with $\alpha$ being the constant coupling parameter of the theory. The spacetime geometry is characterised by the roots of $\bar h$, located at $r_{\pm}=M\pm \sqrt{M^2-\a}$, with the largest one being the event horizon, $r_h=r_+$, whereby $\alpha\leq M^2$. 
Note that, strictly speaking, the black hole inner horizon is ill defined as the scalar becomes imaginary in the interior of the outer horizon{\footnote{This caveat can be remedied by introducing linear time dependence \cite{Charmousis:2021npl}} but we will stick to the static case here for simplicity}. 

Indeed, under the disformal transformation \eqref{disfbh}, the metric functions and scalar field become: 
\be
     h = \bar h\,, \qquad
     f=\frac{\bar h}{1+ 2D(\bar X) \bar X} \,, \qquad
      \f = \bar \f\,. \label{fun-dis}
\ee
The above functions comprise a solution to a beyond Horndeski theory, given by (\ref{disformal2}-\ref{g3d}) and parametrized by $D$.
The new line-element of spacetime therefore reads 
\begin{equation}
   ds^2= - h(r)\,dt^2 + \frac{[1+ 2D(\bar X) \bar X] \,dr^2}{h(r)}+ r^2\,(d\theta^2 + \sin^2 \theta\,d \varphi^2)\,.
   \label{wormhole}
\end{equation}
Taking into account the expression \eqref{phih} for $\bar{\phi}'$ in Horndeski theory, the $\bar X$ function may be written as
\begin{equation}
    \bar X = - \frac{1}{2}\,\bar h \bar{\f}'^2=-\frac{1}{2}\,\frac{(\sqrt{\bar{h}}-1)^2}{r^2}\,,
\end{equation}
and hence the new metric function $f$ is defined from $r=r_h$ to radial infinity. If we choose $D(\bar X)=\beta \bar{X}^\lambda$ where $\lambda>-1$ then asymptotics are not spoiled while if, for example $\beta>0$, $f$ is strictly positive till $r=r_h$. In this case therefore we have a black-hole solution with $f\neq h$ in beyond Horndeski theory. Alternatively choosing $D$ so as to have an additional zero for $f$ at $r=r_0$ with $r_0>r_h$ one can construct a wormhole solution. 

If we define for simplicity the quantity $W=1+2D\bar{X}$, then a disformal transformation of the form
\begin{equation}
\label{W-gen}
    W^{-1}=1+\frac{1}{\lambda^2}\sum_{i=1}^m c_i\left( r_0\sqrt{-2(2n-1)\bar{X}}\right)^i,
\end{equation}
where $(\lambda,\,r_0,\,n,\,c_i)$ are constants, may transform a Horndeski black hole to a beyond Horndeski wormhole. The detailed process was presented in \cite{Bakopoulos:2021liw} where the above transformation with $n=1$, $m=2$, $c_1=-2$ and $c_2=1$ was applied to the Lu-Pang solution \cite{Lu:2020iav}. Let us now apply the above transformation to the Horndeski black hole solution found in Section \ref{first_q}, i.e. the solution described by Eq.(\ref{eqsimplen}). We consider the most simple case\footnote{The choice $c_1\neq 0$ may transform the Lu-Pang black hole ($n=1$) to a wormhole but for $n>1$ adds a small contribution to the $1/r$ term at infinity that spoils the ADM mass.} where $m=2,\,c_1=0$, and $c_2=-1$. For this choice, we find 
\begin{equation}
    \bar{X}=\frac{1-\sqrt{(2n-1)h}}{r\sqrt{(2n-1)h}}, \quad\text{and}\quad W^{-1}=1-\frac{r_0^2(1-\sqrt{(2n-1)h})^2}{\lambda^2r^2}.\label{weqq}
\end{equation}
For the existence of a wormhole, we demand that the function $W^{-1}$ has a root at a value of the radial coordinate $r_0$ larger than $r_h$. The equation  $W^{-1}=0|_{r=r_0}$ then leads to the following condition
\begin{equation}
    h(r_0)=\frac{(1\pm\lambda )^2}{2n-1}. 
\end{equation}
If we demand that $\lambda$ is positive, we must choose the $(-)$ sign in the above equation since, for $r>r_h$, $0<h<1$ and therefore $1-\sqrt{2n-1}<\lambda<1$. For $\lambda=1$, we find $h(r_0)=0$ which means that the throat coincides with the horizon of the black hole ($r_0=r_h$) while for $\lambda=1-\sqrt{2n-1}$ we have $h(r_0)=1$ or equivalently $r_0\rightarrow\infty$, and the throat radius is pushed to infinity. The throat radius of the wormhole satisfies the following equation
\begin{align}
    &(n+1)\,(2 n-1)^n \,r_0^{2 n-1} \left[\,M (4 n-2)+r_0 ((\lambda -2)
   \lambda -2 n+2)\right]\nonumber\\[2mm]
   &+\alpha  \lambda ^{2 n} \,\left[\,\lambda\,(\lambda -2) -2n\, (\lambda -1)+2\right]=0.
\end{align}
While the above equation cannot be solved analytically for $n>1$, as was also the case with (\ref{eqsimplen}), we can find an approximate solution when $\alpha\rightarrow 0$. From Eq. (\ref{eqsimplen}), it is easy to verify that in that limit the metric function $h$ acquires a Schwarzschild form
\begin{equation}
    h(r)=1-\frac{2M}{r}+\mathcal{O}(\alpha)\,,
\end{equation}
while, from $f=h/W$ and (\ref{weqq}), we find
\begin{equation}
    f(r)=\left(1-\frac{2M}{r}\right)\left[1-\frac{r_0^2}{\lambda^2 r^2} \left( 1-\sqrt{(2n-1) \left(1-\frac{2M}{r}\right) }  \right)^2 \, \right]+\mathcal{O}(\alpha).
\end{equation}
From the condition $f(r_0)=0$, and excluding the root that corresponds to the horizon ($r_0=2M$) and the negative root,  we find that the throat radius has the following simple form: 
\begin{equation}
    r_0=\frac{(4n-2)M}{2(n-1)+\lambda(2-\lambda)}+\mathcal{O}(\alpha).
\end{equation}
It thus depends on the mass $M$ and scale parameter $\lambda$, and receives contributions from the coupling constant $\alpha$ of the theory. 

The constructed wormhole solution is a solution to beyond Horndeski theory. The exact form of the coupling functions readily follow by employing the Horndeski functions $G_i$ presented in section 4.1 and the transformation rules \eqref{disformal2}-\eqref{g3d} in conjunction with the expression of the quantity $1+ 2 D \bar X$ given in \eqref{W-gen}.


%

\section{Conclusions}

Horndeski and beyond Horndeski theories are generalised theories of gravity which incorporate standard GR but, at the same time, contain an additional scalar degree of freedom. As a result, they provide an excellent framework for the study of new solutions describing a plethora of compact objects, i.e. black holes, wormholes or regular solutions, with a non-trivial scalar field. The action functional of beyond Horndeski theory is characterised by six arbitrary coupling functions between the scalar field and gravitational quantities, and leads to a set of field equations of increased complexity. Solving analytically these equations is not an easy task with a relatively small number of such solutions having been found so far, especially in the case of flat asymptotics. In the present work, we have considered a general class of scalar-tensor (beyond Horndeski) theories, and formulated different techniques for deriving analytic solutions describing static, compact objects of spherical symmetry. 

To this end, we have introduced four auxiliary functions ($Z$, $Y$, $\A$, $\B$) in terms of which the field equations for the metric functions and scalar field assume a particularly simple form. In the case of shift symmetric, parity preserving theories, parametrised by three coupling functions, ($G_2$, $G_4$, $F_4$), we have demonstrated that the set of field equations is integrable and may, upon appropriate choices, lead to a variety of physically interesting, explicit black-hole solutions. To this end, we have presented several classes of black holes with an asymptotically-flat,  Reissner-Nordstrom behaviour, either homogeneous or non-homogeneous. 
Interestingly, all of these solutions  arise in the context of beyond Horndeski theories containing a canonical kinetic term for the scalar field.

We subsequently turned to the case of non-parity preserving theories and assumed the presence of all six coupling functions, ($G_2$, $G_3$, $G_4$, $G_5$, $F_4$, $F_5$). Although integrability seems to be lost in this case, we developed a filtering technique for a sub-class of theories which allows us to solve the set of field equations upon choosing the form of only one coupling function, $G_4$, and an auxiliary quantity $Q$. The technique was based on the quest for homogeneous solutions but allows also for the emergence of non-homogeneous black holes. To simplify the analysis, we focused on the former case with $f=h$, and presented two indicative sets of analytic black-hole solutions. The first emerges in the context of a theory which is related to the Kaluza-Klein reduction of Lovelock theory, and is a generalisation of the asymptotically-flat solution \cite{Lu:2020iav} to other Horndeski and beyond Horndeski theories. In this case, the spacetime may in fact admit one, two or no horizons depending on the parameters of the solution, with a regular behaviour of the metric function arising at the origin in the last two cases.  The second set of black-hole solutions presented here leads instead to an AdS asymptotic behaviour but exhibits a similar horizon structure as the previous solution; in the cases with two or no horizons, a de Sitter regular core was interestingly found at the origin. 


A different technique was employed in the last part of our work in order to construct non-homogeneous black holes and also a different type of gravitational solutions, namely a wormhole. For the construction of wormhole solutions, the method of disformal transformation was applied to a seed black-hole solution with the disformability function chosen so that the metric function $f$ has a root at a value of the radial coordinate larger than the black-hole horizon. In a previous work \cite{Bakopoulos:2021liw}, such a disformal transformation was applied to a Horndeski solution, i.e. the Lu-Pang black-hole solution \cite{Lu:2020iav},  and a wormhole was constructed with attractive characteristics, such as a regular scalar field and no need for exotic matter.  Here, to illustrate the generality of the construction, we have proposed an alternative form for the disformability function and applied it to the asymptotically-flat families of black-hole solutions found in Section 4. This action leads again to a wormhole solution with a non-trivial, regular scalar field and a throat radius depending on the mass of the seed solution and coupling parameters of the theory.  

Our analysis demonstrates that the beyond Horndeski theory leads to a variety of explicit solutions describing compact objects, black holes and wormholes. These solutions may be derived either by direct integration methods as the general ones we have exposed in sections 3 and 4, or by construction techniques such as the application of a disformal transformation to a previously known, seed solution. Interestingly, construction techniques of the Kerr-Schild type have been shown to lead to non trivial regular black holes but only for the case of DHOST theories \cite{Babichev:2020qpr}, \cite{Baake:2021jzv}. Hence it is not clear if such vacuum solutions exist in beyond Horndeski theories. Our analysis has been general but not exhaustive. For example, although our analysis in section 4 makes some progress in the direction of non-homogeneous black holes without parity symmetry, no explicit solutions were found and it would be interesting to look further into this question as parity breaking theories seem to be particularly interesting departures from GR. As such it would be interesting in particular to find solutions within theories where homogeneous solutions are already known (as depicted in section 4) as uniqueness theorems such as Birkhoff's theorem are not valid in scalar tensor theories and solutions of spherical symmetry, within the same theory, may be in competition in a sense similar to scalarisation for example. 

Another interesting point our analysis has outlined is the existence or not of admissible wormhole solutions in parity symmetric theories. Our general analysis in section 3 shows that the shape function responsible for the presence of a throat, is independent of a (mass) integration constant. This is quite opposite to what happens for parity breaking theories \cite{Bakopoulos:2021liw} where it was shown that the throat disappears whenever mass is set to zero. This seems like a good feature of a wormhole, similar to what happens for most black hole solutions. It is intriguing that for parity symmetric theories, the throat would be independent of an integration constant appearing in the solution and in a certain sense such a throat would be an eternal throat. It would be present for a given theory at a certain (theory dependent) size whatever the mass of the solution. The admissibility of such a throat deserves maybe further study. 

Last but not least, we have considered theories with shift symmetry. This is clearly a mathematically, rather than physically, motivated assumption. Generically there is no reason that such a symmetry would exist in scalar tensor theories and recent considerations, of Kaluza-Klein reduction from Lovelock theory \cite{Charmousis:2014mia} have shown particularly interesting theories which do not have shift symmetry and contain interesting black hole solutions \cite{Fernandes:2021dsb}, \cite{Fernandes:2022zrq}. It would be interesting to study spherically symmetric solutions of such or neighbouring theories with lesser symmetry or none at all. 

\section*{Acknowledgements}
We are very happy to thank Eugeny Babichev and Karim Noui for encouraging and useful discussions throughout the course of this work. The work of N.L. is supported by the doctoral program CDSN ENS Lyon. 
A.B., P.K. and C.C. happily acknowledge networking support by the GWverse COST Action CA16104, “Black holes, gravitational waves and fundamental physics.” C.C. in particular thanks the Department of Physics in the University of Ioannina for hospitality during the course of this work.

\newpage
\appendix
\section*{APPENDIX}
\setcounter{equation}{0}
\renewcommand{\theequation}{\thesection.\arabic{equation}}
\addcontentsline{toc}{section}{APPENDIX\label{app}}

\section{Case of constant kinetic term}

If we assume that $h=f$, the right-hand-side of \eqref{eq1a} vanishes, and we should have either $X'=0$ or $\A=0$. Here, we investigate the first case, i.e. $X'=0$ or equivalently $X=const.$. This easily leads to $\phi' =d/\sqrt{f}$, where $d$ is a constant,
which upon integration, defines the solution for the scalar field. 
The remaining two equations \eqref{scalareqb1} and \eqref{eq2a} should provide the same solution for the metric function $f(r)$, thus they should be of the same form for reasons of consistency. Using the solution for $\phi'$, these take the form
\begin{align} 
 f'\,\left[2 r Z_X + \frac{d}{2 \sqrt{f}}\,(r^2 G_{3X} + G_{5X}) + \sqrt{f} Y_X\right] &= 
G_{2 X} r^2 + 2 G_{4X} -2 r d \sqrt{f} G_{3X}-2 f Z_X, \label{eqf1}\\
 f'\,(2 r Z + 2 \sqrt{f} Y) &= -G_2 r^2 -2 G_4 -2 f Z.\lab{eqf2}
\end{align}
Since for $X=const$, all $G_i$ functions, as well as $Z$ and $Y$, are also constant, the comparison of the above two equations leads to the constraints
\begin{equation}
    Z_X=\lambda Z\,, \quad Y_X=2\lambda Y\,, \quad r^2 G_{3X} + G_{5X}=0\,,
\end{equation}
\begin{equation}
    G_{2X} =-\lambda G_2\,, \quad G_{4X} =-\lambda G_4\,, \quad G_{3X}=0\,.
\end{equation}
The above indeed leads to the trivialisation of the odd functions $G_3$ and $G_5$. From the ghost constraint (2.32),
we then obtain that
\begin{equation}
    F_5=0\,, \quad {\rm or} \quad G_{4}=2X G_{4X}\,.
\end{equation}
The second choice is incompatible with the constraint on $G_4$ derived earlier and thus is discarded. The first choice removes all odd functions from the theory and leads to $Y=0$. Then, integrating \eqref{eqf2}, we obtain a Schwarzschild-(A)dS solution of the form
\begin{equation}
    f(r)=\frac{G_4}{Z} - \frac{2M}{r} - \frac{G_2}{6Z}\,r^2\,.
\end{equation}
In order to restore asymptotic flatness, in the absence of the cosmological constant term, we need to assume that $G4=Z$ which however is also incompatible with the constraints derived earlier unless $X$ itself is trivial.

%

\setcounter{equation}{0}
\section{Additional Classes of Parity-Symmetric Solutions}

The solution \eqref{h_first0} derived in section 3.1 in the context of the parity-symmetric sector of beyond Horndeski theory is only one characteristic example of a broad class of black-hole solutions which may be found in this theory. Insisting on having a canonical kinetic term for the scalar field in the theory, one may consider the following forms of the parity-symmetric coupling functions $G_2$ and $G_4$
\begin{align}
    G_2&=-2\L-\a X +\d X^m, \label{G2_gen}\\[2mm]
    G_4&=\zeta +\b X^n, \label{G4_gen}
\end{align}
where the power coefficients $m$ and $n$ can be in general different
from 2. We will assume again that $Z=\gamma$, and therefore
look for black-hole solutions with $f=h$. Knowing $Z$ and $G_4$ allows us
to determine the remaining parity-symmetric function of the theory, namely
\begin{equation}
    F_4=\frac{\b(1-2n)X^n+\zeta+\gamma}{4 X^2}. \label{F4_gen}
\end{equation}

A large class of physically interesting black-hole solutions may be analytically determined in the case where $n=m$. Then,  (\ref{rel2})  yields the form 
\begin{equation}
   \f'^2=\frac{1}{h(r)}\left|\frac{\alpha  2^{n-1} r^2}{n\,(2 \beta+\delta r^2)}\right|^{\frac{1}{n-1}}, \label{f'_gen}
\end{equation}
for the first derivative of the scalar field. For arbitrary integer, positive, values of $n>1$,   (\ref{rel3}) leads in turn to the solution for the metric function
\begin{align}
    h(r)&=-\frac{\zeta}{\g}+\frac{\Lambda r^2}{3\gamma}+X(r)\left(\frac{2 \beta}{\delta} + r^2\right)\frac{\alpha (n-1)}{6 n \gamma}
    +\frac{\lambda}{2 \gamma r} \nonumber \\
    &- \frac{(n-1) \alpha \beta}{3 n \delta \gamma} \left(\frac{\alpha r^2}{2 n \beta}\right)^{1/n-1}
    {\rm {}_2F_1}\left[\frac{1}{n-1}, \frac{n+1}{2(n-1)}; \frac{3n-1}{2(n-1)}; -\frac{\delta r^2}{2 \beta}\right],
    \label{solh_gen_n}
\end{align}
where $\lambda$ is an integration constant and with ${\rm {}_2F_1}$ we denote the Hypergeometric function ${\rm {}_2F_1}(a,b;c;r)$. For the special case of $n=m=1$, the solution for $h(r)$ may be written as a polynomial and reduces to the Schwarzschild-(A)dS solution for $\zeta=-\gamma$ with a trivial scalar field. For $n=m=2$,   \eqref{solh_gen_n} is expressed in terms of the $\arctan$ function, as also did the solution of section 3.1 where quadratic expressions were similarly assumed for $G_2$ and $G_4$. For $n=m=3$, the metric function is written in terms of radicals whereas, for $n=m>3$, the hypergeometric function cannot be written in terms of elementary functions. 

Nevertheless, for all values of $n=m>1$, the solutions exhibit the same behaviour at asymptotic infinity. There, the solution \eqref{solh_gen_n} reduces to the form 
\begin{equation}
    h(r) \simeq 1- \frac{\Lambda_{eff}\,r^2}{3} -\frac{2 M}{r} + \frac{Q^2}{r^2} + \cdots, \label{h_gen_asym}
\end{equation}
thus describing an (A)dS-Reissner-Nordstrom solution
under the identifications
\begin{equation}
    \beta=-\frac{n \delta (\zeta+\gamma)}{\alpha}\left(\frac{n \delta}{\alpha}\right)^{1/n-1}, 
    \end{equation}
\begin{equation}    
     M = -\frac{\lambda}{4\gamma} - \frac{\alpha \beta (n-1)}{3 n\delta \gamma}
    \left(\frac{\alpha}{n \delta}\right)^{1/n-1} \sqrt{\frac{2\pi \beta}{\delta}}
    \,\frac{\Gamma\left[\frac{3n-1}{2 (n-1)}\right]}{\Gamma\left[\frac{1}{n-1}\right]}\,.
    \end{equation}
The first relation ensures the asymptotic flatness of spacetime, in the absence of a cosmological constant, while the second determines the mass of the black hole in terms of the integration constant $C$ as well as the remaining parameters of the theory. The effective cosmological constant turns out to be given by the combination 
\begin{equation}
   \Lambda_{eff} \equiv -\frac{\Lambda}{\gamma} -\frac{\alpha (n-1)}{2 n\gamma}\left(\frac{\alpha}{n \delta}\right)^{1/n-1},
\end{equation}
and is sourced by both the bare term $\Lambda$ and the canonical kinetic term of the scalar field appearing in the expression of $G_2$. Therefore, a (beyond) Horndeski theory with a canonical kinetic term for the scalar field generically leads to a non-asymptotic flat spacetime. Asymptotically-flat black-hole solutions may of course arise if the parameters of the theory are chosen so that the right-hand-side of the above equation vanishes. Finally, the solution is ``dressed'' with a tidal charge given by
\begin{equation}
    Q^2 \equiv -\frac{\alpha \beta^2}{\gamma \delta^2 (n-1)}
    \left(\frac{\alpha}{n \delta}\right)^{1/n-1},
\end{equation}
which is also an inherent feature of our solutions for generic values of the parameters of the theory.

Black-hole solutions with either an exact or an asymptotic (A)dS-Reissner-Nordstrom behaviour arise also in the case where $m \neq n$ in   \eqref{G2_gen}-\eqref{G4_gen}. For instance, if we keep $m=2$ as above but instead set $n=1$, so that $G_4$ is linear in $X$, \eqref{rel2} gives the result
\begin{equation}
    \f'^2=\frac{2\b -r^2\a}{\d r^2 h(r)},
\end{equation}
while   \eqref{rel3} leads to the following solution for the metric function
\begin{equation}
    h=1-\frac{2M}{r}+\frac{2\d(\g+\zeta)^2}{\g \a r^2} +
    \Bigl(\Lambda+\frac{\alpha^2}{8\delta}\Bigr) \,\frac{r^2}{3\gamma}. 
\end{equation}
The above is an exact (A)dS-Reissner-Nordstrom solution where we have made the choice $\b =-2\d(\g+\zeta)/\a$. It is a solution of the beyond Horndeski theory since $F_4$ is still given by  \eqref{F4_gen} with $n=1$. For $m=2$ and $n=3$, we also obtain a black-hole solution with a non-trivial $\phi$ and a metric function expressed in terms of polynomials and radicals; however, for brevity, we refrain from showing here its explicit from as its asymptotic (A)dS-Reissner-Nordstrom behaviour is similar to the one presented in   \eqref{h_gen_asym}. 

A particularly simple configuration arises if we set $\delta=0$ in   \eqref{G2_gen}. Then, the coupling function $G_2$ is given only in terms of the bare cosmological constant $\Lambda$ and a canonical kinetic term. The profile of the scalar field is still given by   \eqref{f'_gen} with $\delta=0$. It is easy to see that $\phi'$ remains regular everywhere only for $n<2$. As noted earlier, the case $n=1$ leads to the pure Schwarzschild-(A)dS solution, and the same holds for the case with $n=0$. The case with $n=1/2$ does lead to a black-hole solution with a non-trivial scalar field described by the equation
\begin{equation}
     \phi'^2=\frac{2 \beta^2}{\alpha^2 r^4 h}\,,
\end{equation}
and a metric function given by the expression
\begin{equation}
    h=-\frac{\zeta}{\g}-\frac{2M}{r} +\frac{\L r^2}{3\g} +\frac{\beta^2}{2 \alpha \gamma}\,\frac{1}{r^2}.
    \label{h_C}
\end{equation}
Upon choosing $\zeta=-\gamma$, asymptotic flatness is restored and the solution describes again an (A)dS-Reissner-Nordstrom spacetime. However, this is a solution to Horndeski theory since the choices $n=1/2$ and  $\zeta=-\gamma$ trivialise the function $F_4$ according to   \eqref{F4_gen}. This solution has been studied in detail in \cite{Babichev:2017guv}. In order to keep $F_4$ in the theory, though, we choose instead $n=1/3$. Then,   (\ref{rel2}) gives
\begin{equation}
     \phi'^2=
    \left(\frac{2}{3}\,\frac{\beta}{\alpha}\right)^{3/2}\,\frac{1}{r^3 h}\,,
\end{equation}
and leads to a vanishing $\phi'$ at infinity. The last equation,   (\ref{rel3}), yields for the metric function the exact expression
\begin{equation}
    h=1-\frac{2M}{r} +\frac{\L r^2}{3\g} -\frac{2\beta}{3 \gamma}\,\sqrt{\frac{\beta}{\alpha}}\,\frac{\ln r}{r},
    \label{h_C2}
\end{equation}
where we have again set $\zeta=-\gamma$. The above expression describes a spacetime which is asymptotically (A)dS with the cosmological constant coinciding with the bare one $\Lambda$. The Schwarzschild term proportional to the mass $M$ of the black hole is now supplemented by an unusual logarithmic term which although vanishes at infinity it does so slower than the mass term. This solution arises in the context of beyond Horndeski theory since our choices leave a non-vanishing $F_4$ function, i.e. $F_4=\b/(12 X^{5/3})$.


\setcounter{equation}{0}
\section{Parity symmetry breaking with a beyond Horndeski term}

We will now study a case which stands between parity-symmetric and parity-breaking beyond Horndeski theories. To this end, we will keep the assumption of a parity-symmetric Horndeski sector and thus set $G_3 = G_5=0$. Then, the no-ghost condition \eqref{BH} leads to $0 = F_5\left(G_4-2XG_{4X}\right)$. If one sets $G_4(X) = \sqrt{-\beta X}$, where $\beta >0$, the no-ghost condition is satisfied without taking $F_5=0$, that is without killing the non-parity symmetric part of the beyond Horndeski sector. However, for simplicity, we may consider $F_4=0$. Then, from the definitions \eqref{zdef}-\eqref{ydef}, we readily find
\be
Z(X) = 0\,, \qquad 
Y(X) = 3\left(-2X\right)^{5/2}F_5\,,
\ee
whereas the definitions \eqref{Adef}-\eqref{Bdef} lead to 
\be
\label{ABmixed}
\mathcal{A} = 2\sqrt{f}Y_X \,, \qquad 
\mathcal{B} = \sqrt{f}Y \,.
\ee
Then, the field equation (\ref{eq1a}) integrates to 
\begin{equation}
    f=\frac{\gamma h}{Y}\,,
\end{equation}
where $\gamma$ is an integration constant.
Here, we will look for homogeneous black-hole solutions with $f=h$. Then, the aforementioned equation leads to $Y=\gamma$, which completely determines $F_5$, namely $F_5=\frac{\gamma}{3\left(-2X\right)^{5/2}}$. For $Y=\gamma$, we also obtain $\mathcal{A}=0$ and $\mathcal{B} = \sqrt{f} \gamma$ from \eqref{ABmixed}. Then, the remaining equations of motion \eqref{scalareqb1}-\eqref{eq2a} become:
\begin{align}
0 = G_{2X}r^2-\frac{\beta}{\sqrt{-\beta X}}\,, \label{eq1} \\
2\gamma \sqrt{f} f'= -G_2r^2 - 2\sqrt{-\beta X}\,. \label{eq2}
\end{align}
Setting the form of the last coupling function $G_2$ will fix the theory, and provide the solution to the scalar field and metric function via \eqref{eq1}-\eqref{eq2}, respectively. 

We will study in detail the case where 
$G_2(X)=\eta^2 X + p \sqrt{-X}$, where $(\eta, p)$ are constant coefficients. In this form of $G_2$, the canonical kinetic term of the scalar field is supplemented by an additional $\sqrt{-X}$ term. Equation (\ref{eq1}) now 
determines $X$
\begin{equation}
   X = -\frac{\left(2 \sqrt{\beta }+p r^2\right)^2}{4 \eta ^4 r^4}\,,
\end{equation}
and the scalar field itself through the definition $X=-f \phi'^2/2$. Using the forms of $G_2$ and $X$, the latter equation (\ref{eq2}) takes the form 
\begin{equation}
    2 \gamma  \sqrt{f} f'+\frac{4 \beta + p^2 r^4+4 \sqrt{\beta } p r^2}{4 \eta
   ^2 r^2}=0\,.
\end{equation}
The above integrates to 
\begin{equation}
 f(r)=   \frac{1}{4} \left(\frac{12 \beta -p^2 r^4- 12
   \sqrt{\beta } p r^2+24 \gamma  \eta ^2 \lambda  r}{2\gamma  \eta ^2
   r}\right)^{2/3}\,,
\end{equation}
where $\lambda$ is an integration constant. The expansion at infinity is 
\begin{align}
   f(r) =\left(-\frac{p^2}{16\gamma  \eta^2}\right)^{2/3} r^2 +\frac{\sqrt{\beta } \left(-\frac{\sqrt{2}\,p^2}{\gamma  \eta
   ^2}\right)^{2/3}}{p} + \frac{2\lambda}{\left(-\frac{p^2}{2\gamma 
   \eta ^2}\right)^{1/3}r} 
   -\frac{3\beta  \left(-\frac{\sqrt{2}\,p^2}{\gamma  \eta
   ^2}\right)^{2/3}}{p^2 r^2}+\mathcal{O}\left( \frac{1}{r^3} \right)
\end{align}
which describes an (A)dS-Reissner-Nordstrom background for $p>0$ and $\gamma <0$. Note that the ADM mass 
\begin{equation}
  M=  2\lambda \left(\frac{2|\gamma|
   \eta ^2}{p^2}\right)^{1/3}
\end{equation}
depends on the integration constant $\lambda$, and is therefore a free parameter of the solution, while the effective cosmological constant and tidal charge are fully determined by the coupling parameters ($\beta,\eta, p,\gamma$).


\bibliographystyle{utphys}
\bibliography{Bibliography}

\end{document}